\documentclass{article}[15pt]      

\usepackage{amsmath,epsfig,color,calc,rotating,graphics}
\usepackage{ulem}
\usepackage{lineno}

\usepackage{float}

\usepackage{natbib}
\bibpunct{(}{)}{;}{a}{}{,}

\begin{document}

\title{An  Extensive Study of Two-Node McCulloch-Pitts Networks 
\vspace{0.2in}
\author{
Wentian Li$^{1,2}$, 
Astero Provata$^3$,
Thomas  MacCarthy$^1$ \footnote{Prof. Thomas MacCarthy (1970-2023) passed away during the preparation of the manuscript.}
\\
{\small  1. Department of Applied Mathematics and Statistics, Stony Brook University, Stony Brook, NY, USA}\\
{\small  2. The Robert S. Boas Center for Genomics and Human Genetics}\\
{\small  The Feinstein Institutes for Medical Research, Northwell Health, Manhasset, NY, USA}\\
{\small \sl 3. 
Institute of Nanoscience and Nanotechnology }\\
{\small \sl National Center for Scientific Research, ``Demokritos", 15341 Athens, Greece}\\
}
\date{\today}      
}  
\maketitle                   
\markboth{\sl Li et al.}{\sl Li et al.}

\normalsize

\begin{center}
{\bf ABSTRACT}
\end{center}

Networks with two nodes are previously grouped into either two
classes (mutually interactive, master-slave) or five classes (mutualism,
competition, predator-prey, commensalism, amensalism). By allowing
self-loops, the number of signed regulatory graphs increases to 39. We
provide a complete summary of dynamical behaviors of the 39
two-node McCulloch-Pitts models when the link weights are
constrained to three values [$-1$,0,$+1$] and Boolean node variables. 
Depending on whether the Boolean values
are [$-1,1$] (bipolar) or [0,1] (binary), we show that the dynamics could
also be different with the same signed regulatory graphs. 
We demonstrate that slight variations in the McCulloch-Pitts model 
(called variants) may lead to fundamentally different dynamics.
We study the full model space and three kinds of robustness or stability:
a) of a rule against parameter change on its overall dynamics,
b) for a given state against parameter change on its final state, 
and c) against an initial state change on its final state.
All these stability properties are loosely related to a model's limiting dynamics,
with the fixed-point rules to be more stable in the first two types of robustness,
but less stable in the third robustness type.
These analyses pave the way towards a better understanding of a
minimum complex system.

\vspace{0.2in}

\newpage

\large

\section{Introduction}

\indent

The McCulloch-Pitts (MP) neuron was first proposed in the literature
as a simple neuron firing model \citep{mcculloch}.
A specific MP neuron receives inputs
that usually come from connections with other MP neurons as well as
from external controls or even from random environmental noise.
A weighted sum operation of the inputs is then performed,
 followed by a threshold operation and the output is propagated
to other neurons in the network.  The additive
contribution from multiple inputs is the most often used functional 
form in modelling, whenever details of the specific joint action 
(or epistasis) are unknown. Examples include the polygenic risk
score used in the study of complex disease genetics \citep{purcell} 
which was predated
by RA Fisher’s polygenic model \citep{fisher1918,visscher}.  
We now understand that, although a single MP neuron may be a simple
integrator with a threshold, networks of cooperating MPs may produce
intricate unexpected behavior.
The aim of the present study is to explore extensively the minimal level
of complexity arising in networks consisting of only two nodes and show that
even this small network may lead to nontrivial dynamical patterns,
giving some first glimpses of the dynamics to be expected in larger networks.

While integrating contributions from neurons in a network, 
the threshold function is an effective way of handling saturation \citep{goutelle}.
It is normally extremely nonlinear, though could be piecewise linear. 
There could be other more sophisticated nonlinear functions/mappings learned from the data 
\citep{otwin}, 
but the threshold function is the simplest. The network with McCulloch-Pitts gates to propagate
signals from one layer of nodes to another is also known as perceptron \citep{perceptron}, 
 or threshold logic unit networks \citep{TLU}, threshold-element networks 
\citep{amari}, and threshold networks \citep{goles}. The combination of weighted sum
and a nonlinear filter function (logistic function) is also a familiar form in statistics in the case
of logistic regression \citep{GLM}.

\begin{figure}[H]
\begin{center}
  \begin{turn}{-90}
  \end{turn}
   \includegraphics[width=0.8\textwidth]{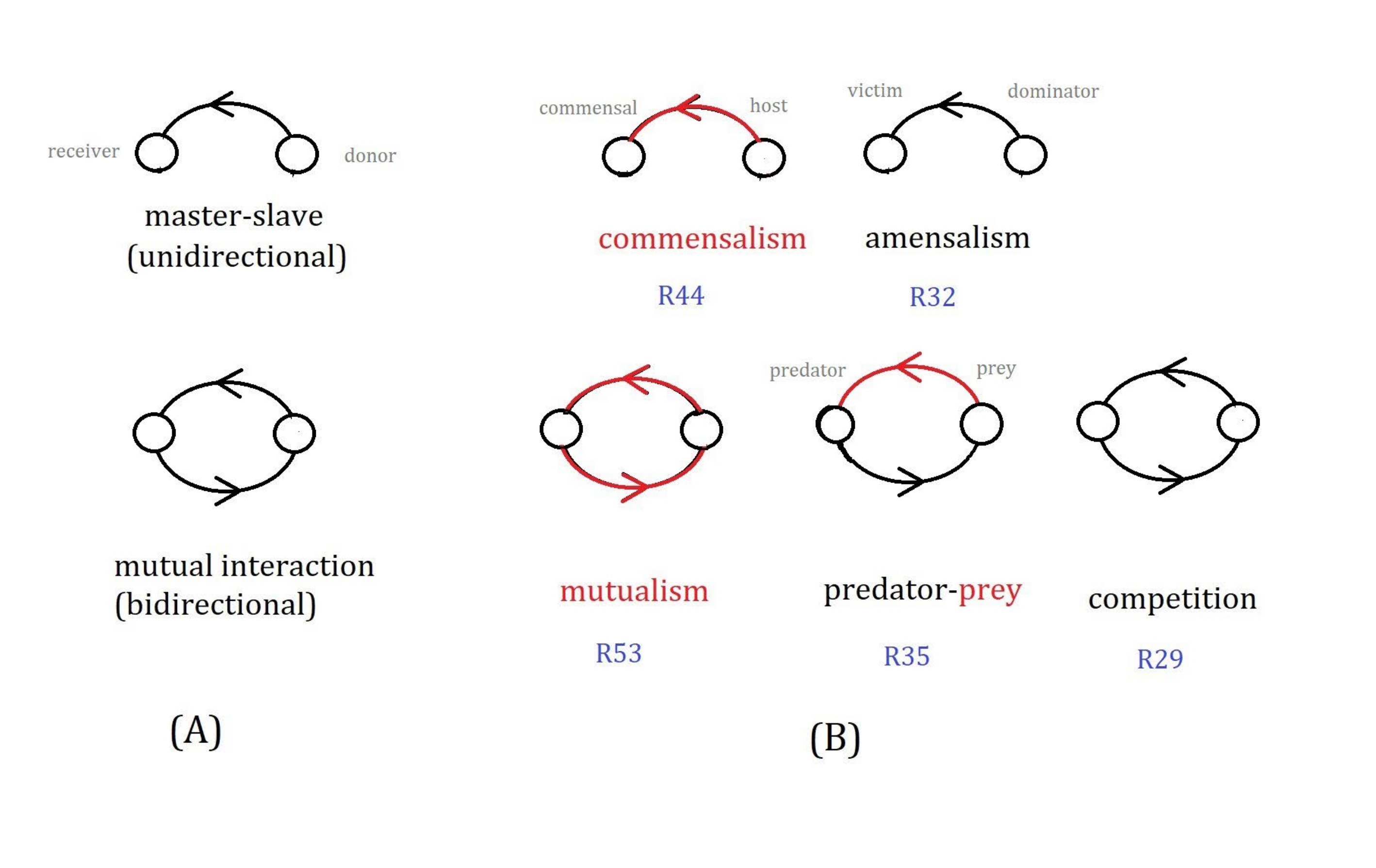}
\end{center}
\caption{
\label{fig1}
(A) The two basic types for two-node interaction:
master-slave(unidirectional) and mutual-interaction (bidirectional).
(B) The five basic types for two-node interaction when 
positive (red) or negative (black) signs are added to the arrows:
commensalism (unidirectional positive), amensalism (unidirectional negative), 
mutualism (two positives links), predator-prey (one positive and one 
negative links), and  competition (two negatives links).
}
\end{figure}

To simplify further the 2-node system, in the present study the state variable 
on the nodes is confined to be Boolean or two-valued.  The threshold function 
is very appropriate for confining the weighted output to Boolean type.
Regarding the interactions between the two units, there are only two modes 
of operation: a directed link from one node to another without an arrow in 
return (master-slave or unidirectional), and, both nodes have a directed link to 
the other node (mutual interaction or bidirectional). 
A schematic representation of the master-slave and 
mutual interactions is presented in Fig. \ref{fig1}(A).

When a specific MP function is used, it becomes clearer that the master-slave and mutual
interactive classification is not enough. The directed link from one node to another can be
positive (excitatory) or negative (inhibitory), besides being zero (absent). In the field of
ecology, it has been long known that there are five types of interaction between two species:
commensalism (positive master-slave), amensalism (negative master-slave), mutualism (two
positive links) \citep{boucher,fath,henderson},
competition (two negative links), and predator-prey (one positive and one negative links) 
\citep{wangersky}
(see chapter 7 of \citep{odum}, chapter 9 of \citep{williamson}, and chapter 2 of \citep{may}).
We avoid the name symbiosis because it may have different definitions by 
different authors \citep{martin}.  
These five classes are provided in Fig. \ref{fig1}(B).

These five ecological classes of interactions do not include
intra-species activities. The Allee effect describes a positive
correlation between population size (up to certain point)
and the fitness of the population \citep{allee31}. Autocatalytic
chemical reactions have been proposed as a mechanism for origin of life
\citep{eigen}. A gene might be
regulated by its own gene product during transcription. All these would
mean a link, either positive or negative, from a node to itself.
Inclusion of self-link will increase the number of two-node networks
from 5 to 39 (see Tables \ref{table1} and \ref{table2-39rules}).

\begin{figure}[H]
\begin{center}
  \begin{turn}{-90}
  \end{turn}
   \includegraphics[width=1.1\textwidth]{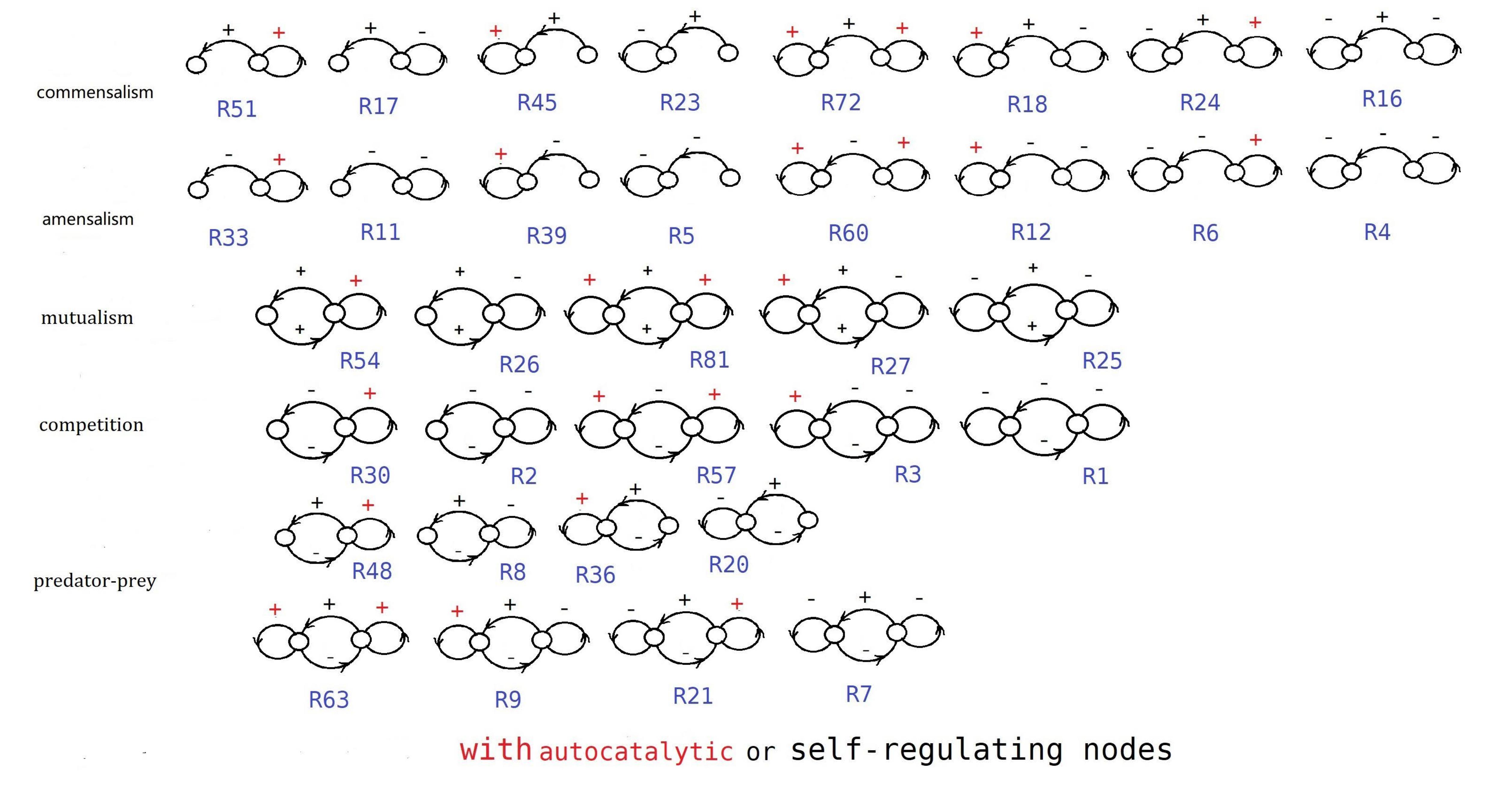}
\end{center}
\caption{
\label{fig2}
The five basic types in Fig.\ref{fig1} are expanded to 39
models (34 here and 5 in Fig.\ref{fig1}) by including self links.
Positive self links are called autocatalysis (autocatalytic),
and negative self links self-regulation (self-regulating).
}
\end{figure}

To keep the MP networks as simple as possible, besides restricting the node 
number to two nodes and the network state variables to two possible values, 
we also restrict the link strength to three possible values only (positive one, 
negative one, and zero), and restrict the threshold value to be zero 
(or close to zero). 
Because there are four allowed links between two nodes,
including those to itself, each having three possible values,
there are total $3^4=81$ two-node
MP network models. The difference between 81 and 39 is due to 
the facts that some of the 81 models are actually disjoint one-node models
(see Supplement Material,, item 1).
and some models are equivalent to each other by exchanging two nodes. 
The graphs of the expanded 39-5=34 equivalent models are drawn in Fig \ref{fig2}.
These graphs with defined (positive or negative) parameter
values are called ``signed regulatory graphs (or structures)" in this paper.

The 81 MP models, also known as McCulloch-Pitts Networks (MPN), 
form a (unfolded) ``rule space"
(see Supplement Material, item 2 for a discussion
on different spaces), while merging
equivalent models into one leads to a ``folded" rule space".
From our experience with other rule spaces, such as the 
cellular automata \citep{wolfram-nature,wolfram-book}
rule space \citep{wli-ca,wli-D}, non-local cellular automata
rule space \citep{wli-nl},
the space of two-locus disease models \citep{wli-2l}, and the space
of two-person games \citep{rapoport,marris}, the use of unfolded space
is more natural.

There are two main goals in studying these minimal  MPNs.  The 
first is on a possible link between signed regulatory graph and dynamics.  
Rene Thomas had proposed two types of basic mode in feedback circuits 
\citep{thomas78,thomas81,thomas-pos,thomas-book,thomas01,memory19}:
if there is a path which is overall negative (product of signs along the path
being negative), the dynamics is more likely cyclic;
if there is an overall positive loop, there could be multiple unstable limiting
dynamics. But one needs to be careful in equating his
logical structure with our signed regulatory structure based on MP model.

Towards this, we noticed that the two node state values can either be 
$[-1, 1]$, termed ``bipolar", or (0,1), termed ``binary". Models using 
bipolar values and those using binary values are not equivalent, because 
the transformation between them would lead to a non-zero threshold value. 
Also, the mapping at the exact threshold point
is not definitely specified \citep{snoussi93}, and choosing one
output out of two possible values is similar to increase or decrease 
the threshold slightly from the gap.  These lead to many variants of a MPN model, 
even if the regulatory graph of the model is identical.  Different variants 
will be introduced in detail in the Models and Method section. 
We will show that the dynamics not only possibly depends on 
regulatory structure but also on variants. 

The second goal relates to the stability of the dynamics under minimal 
model perturbations.
We investigate three types of stability: 
(a) the limiting dynamics of a rule against changes in the model parameter;
(b) the limiting state against changes in the model parameter;
and
(c) the limiting state against changes in the initial state.
In the model space, we expect the existence of regions where rules 
behave similarly while the boundaries between these regions
are intricate reminding those at the ``edge of chaos" \citep{packard,langton}. 
With only four states (see Supplement Material, item 2 on state space), 
there will be no chaotic dynamics, but only cyclic dynamics with a maximum cycle length. 
Therefore, we aim at finding the edge between simple fixed points
and high cycles.

This manuscript is organized as follows:
The Models and Methods section introduces the mapping and notation
of the MPN model studied here (Sec.~\ref{Msec:definitions}). 
We then in Sec.~\ref{Msec:variants}  introduce six variants
per model, depending on whether the state variable
is bipolar or binary, and what the choice at the threshold line is. 
Section ~\ref{Msec:spectrum} introduces
the method we used to determine the limiting dynamics, the spectral method, 
as well as the shorthand notations for different types of dynamical behavior.
Sec.~\ref{Msec:program} covers the programs used.

The result section contains 4 subsections:
Sec.~\ref{Rsec:difvar} shows dynamics of different variants;
Sec.~\ref{Rsec:thomas} is an attempt to address the question on  whether 
limiting dynamics is related to the model's regulatory graph. 
Sec.~\ref{Rsec:rulespace} is about robustness of a model's limiting 
dynamics with respect to perturbation on the link parameter, using the 
first variant (V1) as an example;
Sec.~\ref{Rsec:rulespace2} is about stability of limiting attractor 
with respect to perturbation of the link parameter,
as well as with respect to perturbation in the initial state,
using the fourth variant (V4) as an example. 

The Discussion section briefly addresses 
relationship between the MPNs and Hopfield, Kauffman, and Wagner networks;
the potential relation
between MPNs and recurrent neural networks; 
asynchronous updating; allowing mapping output at the
threshold line to be a third value, which can be done by replacing
the sign/step discontinuous function with the continuous sigmoid
function; some situations in MPN
not addressed in this paper; and possible future works.

\section{Models and Methods}

\indent

In the present section, we describe in detail the main MPN models
with all variants, together with the methods used for the analysis
of the 2-node MPN network dynamics.

\subsection{Notations for $N=2$ McCulloch-Pitts networks}
\label{Msec:definitions}

\indent

To relate the terminologies between general networks and
gene regulatory networks \citep{smolen,thakar24}, we use genes
and nodes interchangeably.
See Supplement Material, item 2 for discussion
on gene/node space and other spaces.

Denote $x_i^t$ the expression level of gene-$i$ (transcription factor $i$)
at time $t$, and weight $w_{i \leftarrow j}$ the contribution from gene-$j$ to gene-$i$.
With two genes ($N=2$), we denote $x_1=x,\> x_2=y$,  which can only take two values:
$-1$ and 1 (these are called $[-1,1]$ models \citep{huerta}, or bipolar models),
or alternatively, 0 and 1 (these are called [0,1] models \citep{huerta},
or binary models).
We denote $w_{11}=a, \> w_{12}=b, \> w_{21}=c, \> w_{22}=d$, which can only take three
possible values: $-1$ (suppression/inhibition), 0 (no link), 1 (activation/excitation).
A MPN rule is of the form:
\begin{eqnarray}
\label{eq-mpn}
 x^{t+1} &=& f( a x^t +b y^t) \nonumber \\
 y^{t+1} &=& f( c x^t +d y^t)
\end{eqnarray}
where $f()$ is a threshold function, which is Sign function for
bipolar models, and Step function for binary models.

A $N=2$ network rule is completely specified by the $(a,b,c,d)$ parameters,
and with three possible values in each parameter, there are $3^4=81$ rules/models
(unfolded rule space).
We define a rule/model number $R$, with $R=1$ for ($a=b=c=d=-1$),
$R=2$ for ($a=b=c=-1, d=0), \cdots, R=81$ for $(a=b=c=d=1)$;
or $R= 27 \times (a+1) + 9 \times (b+1)+ 3 \times (c+1) +(d+1) +1$.
When the sign of $(a,b,c,d)$ parameters, including
presence of absence,  are fixed, we have a regulatory graph,
i.e., one of the graphs in Fig.\ref{fig1} or \ref{fig2},
based on the presence/absence and the sign of $(a,b,c,d)$.
Here, because we only have one parameter value for one sign
(e.g. no $a=0.5$ or $a=2$), once the regulatory structure is fixed,
the model/rule is also fixed, and vice versa. 
Variants, on top of a model, will be discussed in 
subsection \ref{Msec:variants}.

\subsection{Variants of McCulloch-Pitts networks}
\label{Msec:variants}
\indent

Fixing the number of genes, the presence or absence of links, the arrow direction,
and activation/inhibition nature of the arrow, still do not completely 
fix  the mapping function in Eq.(\ref{eq-mpn}). 
Many seemingly minor variations can actually affect the dynamics,
including the choice of two state values, and the treatment at the threshold
point of the filtering function, also known as activation function.
The reasons for this are because we limit our MPN models without threshold point
as a free parameter (i.e., $x^{t+1}=f(ax^t+by^t-g)$ with $g=0$), 
and because the discretization leaves 
the mapping output a discontinuous gap  at the threshold point. 
 We list six variants of a MPN rule as follows.

{\bf 1. The ``bipolar as it is" variant (V1) of MPN:}
We call the following dynamical system the ``bipolar as it is" version of MPN:
\begin{eqnarray}
\label{m1p1-asitis}
x^{t+1}  &= & \left\{
\begin{array}{cl}
\text{Sign} ( ax^t +b y^t )  &
 \text{if  } ax^t +by^t  \ne 0  \\
x^t & \text{if  } ax^t+by^t =0
\end{array}
\right.  \nonumber \\
y^{t+1}  &= & \left\{
\begin{array}{cl}
\text{Sign} ( cx^t +d y^t )  &
 \text{if  } cx^t +dy^t  \ne 0  \\
y^t & \text{if  } cx^t+dy^t =0
\end{array}
\right.
\end{eqnarray}
Note that the same ``as it is" is applied to both $x$ and $y$.

{\bf 2. The ``bipolar positive" variant (V2) of MPN:}
In this version, the treatment when the weighted sum is zero is different:
\begin{eqnarray}
\label{m1p1-pos}
x^{t+1}  &= & \left\{
\begin{array}{cl}
\text{Sign} ( ax^t +b y^t )  &
 \text{if  } ax^t +by^t  \ne 0  \\
1 & \text{if  } ax^t+by^t =0
\end{array}
\right.  \nonumber \\
y^{t+1}  &= & \left\{
\begin{array}{cl}
\text{Sign} ( cx^t +d y^t )  &
 \text{if  } cx^t +dy^t  \ne 0  \\
1 & \text{if  } cx^t+dy^t =0
\end{array}
\right.
\end{eqnarray}
The difference between Eq.(\ref{m1p1-pos}) and Eq.(\ref{m1p1-asitis}) is only
in the second lines.
This version was used in (e.g.) \citep{greil,pinho}.
Some studies may define implicitly that
$x_i^{t+1}=0$ when $\sum_j w_{i \leftarrow  j} x_j^t =0$
\citep{wagner96,wagner-martin07-pnas}, which would lead to a state value
of 0 besides $-1$ and 1 (to be discussed in the Discussion section
on sigmoid function).
However, it was not considered a serious issue when the weights are
sampled from a normal distribution,
as the chance for $\sum_j w_{i \leftarrow  j} x_j^t =0$ is practically zero.
For our discretized and simplest MPNs, however, it is important
to distinguish version V2, Eq.(\ref{m1p1-pos}), from version V1, Eq.(\ref{m1p1-asitis}).

There is an alternative version of Eq.(\ref{m1p1-pos}). We may add a small
threshold value $0 < \epsilon < 1$:
\begin{eqnarray}
\label{eq-v2-neg-epsi}
x^{t+1} &=&  Sign (a x^t +b y^t +\epsilon) \nonumber \\
y^{t+1} &=&  Sign (c x^t +d y^t +\epsilon)
\end{eqnarray}
This way, (e.g.) if $a x^t + by^t=0$, $x^{t+1}=1$.

{\bf 3. The ``bipolar  negative" variant (V3) of MPN:}
One can also define the ``bipolar negative" version by assuming
$x_i^{t+1}=-1$ and $y^{t+1}=-1$  at the threshold point:
\begin{eqnarray}
\label{m1p1-neg}
x^{t+1}  &= & \left\{
\begin{array}{cl}
\text{Sign} ( ax^t +b y^t )  &
 \text{if  } ax^t +by^t  \ne 0  \\
-1 & \text{if  } ax^t+by^t =0
\end{array}
\right.  \nonumber \\
y^{t+1}  &= & \left\{
\begin{array}{cl}
\text{Sign} ( cx^t +d y^t )  &
 \text{if  } cx^t +dy^t  \ne 0  \\
-1 & \text{if  } cx^t+dy^t =0
\end{array}
\right.
\end{eqnarray}
Again, there is an equivalent version by subtracting a small positive
threshold $\epsilon$ in the equation:
\begin{eqnarray}
\label{eq-v3-pos-epsi}
x^{t+1} &=&  Sign (a x^t +b y^t -\epsilon) \nonumber \\
y^{t+1} &=&  Sign (c x^t +d y^t -\epsilon)
\end{eqnarray}

The V3 variant (bipolar negative) has the identical dynamics as V2,
because after the variable transformation $x'=-x, y'=-y$, 
$ -x'^{t+1} = x^{t+1} = Sign(-ax'^t -by'^t - \epsilon)$, or
$x'^{t+1}= Sign(ax'^t +by'^y + \epsilon)$, which is the same as the V2 rule.

{\bf 4. The ``binary  as it is" variant (V4) of MPN:}
\begin{eqnarray}
\label{eq-v4}
x^{t+1}  &= & \left\{
\begin{array}{cl}
\text{Step} ( ax^t +b y^t )  &
 \text{if  } ax^t +by^t  \ne 0  \\
x^t & \text{if  } ax^t+by^t =0
\end{array}
\right.  \nonumber \\
y^{t+1}  &= & \left\{
\begin{array}{cl}
\text{Step} ( cx^t +d y^t )  &
 \text{if  } cx^t +dy^t  \ne 0  \\
y^t & \text{if  } cx^t+dy^t =0
\end{array}
\right.
\end{eqnarray}
which is similar to Eq.(\ref{m1p1-asitis}) 
except the step function is used instead of sign function,
and the state variable values  $x_i \in [0,1]$ instead of $[-1,1]$.
Of course, the step function is essentially a sign function
with only a difference of the step size (a factor of 1/2).
This version was used in (e.g.) \citep{tang04}.
The names binary vs bipolar used here were
[0,1]-model vs $[-1,1]$-model  in \citep{huerta}.

{\bf 5. The ``binary positive" variant (V5) of MPN:}
The same equation  Eq.(\ref{m1p1-pos}) is used, but the state value
$x_i \in [0,1]$, and sign function is replaced by the step function.

{\bf 6. The ``binary  negative" variant (V6) of MPN:}
The same equation  Eq.(\ref{m1p1-neg}) is used, but the state value
$x_i \in (0,1)$, and sign function is replaced by the step function.

Note that even with those variants listed here,
we have not yet exhausted all possible variants of MPNs,
including:
(1) Another version of MPN which is closer to a 
differential equation, while keeping the node variable non-negative.
We call this variant ``binary difference" or V7
(see Supplement Material, item 3). Interestingly, this
version is equivalent to V4 (binary as it is). 
(2) Asynchronous updating.  At each time step, only one node
(e.g. node-1 or $x$) is updated, and the updating of the second 
node (e.g. node-2 or $y$) uses the new
value at node-1 as input. 
(3) One updating variant for $x$ and another for $y$, choosing among, e.g.,
V4, V5, V6. This is equivalent of using different thresholds for $x$ and $y$ node.
(4) The use of sigmoid function as
activation function which is widely used in artificial neural networks. 
This would solve the problem at the threshold point, because
the discontinuous gap is gone. But at the same time it creates 
a new problem. For bipolar models for example, 
a third state value, the zero, may appear, making it a three-level
model (see the Discussion section).

\subsection{Spectrum property of state space}
\label{Msec:spectrum}

\indent

\begin{figure}[H]
\begin{center}
  \begin{turn}{-90}
  \end{turn}
   \includegraphics[width=0.7\textwidth]{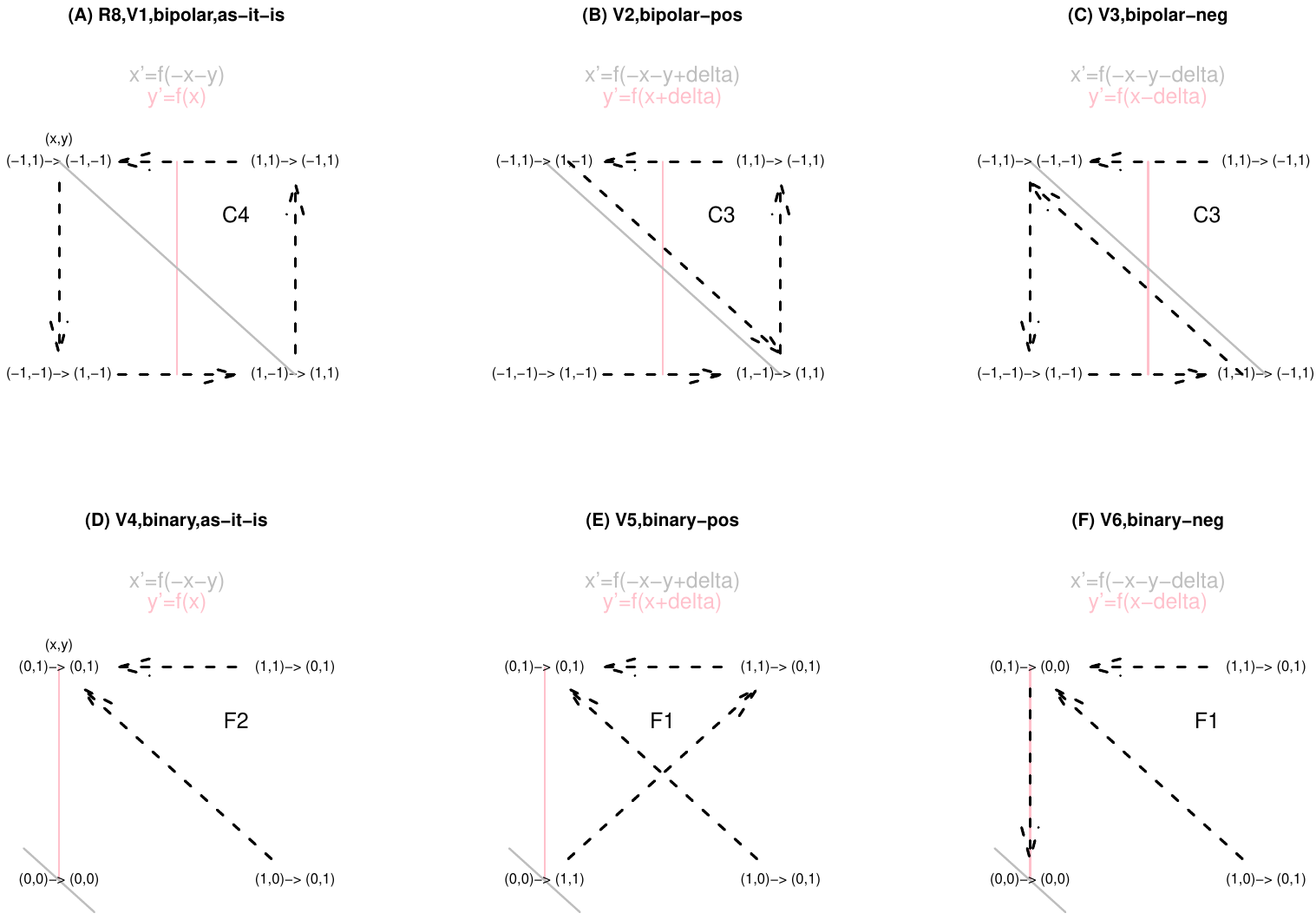}
\end{center}
\caption{
\label{fig3-R8}
Illustration that given the same regulatory graph, the limiting dynamics
can depend on the variants. The rule R8 is used,
with $a=b=-1$, $c=1$, $d=0$.
The $x$ axis represents the node-1, and $y$ for node-2.
 grey/pink line represents the threshold line for $x$/$y$ of 
Eq.(\ref{eq-mpn}). The dashed lines are the state transition (a state
without a dashed line means that it has a transition to itself).
(A) bipolar as it is (V1), 4-cycle;
(B) bipolar positive (V2), 3-cycle;
(C) bipolar negative (V3), 3-cycle;
(D) binary as it is (V4), two fixed-points;
(E) binary positive (V5), one fixed point;
and
(F) binary negative (V6), one fixed-point.
}
\end{figure}

The dynamics of a network rule can also be determined by the one-step
mapping in the state space. Take the R8 in V1 variant 
for example (Fig.\ref{fig3-R8}(A)), 
$S_0=(-1,-1) \rightarrow S_2=(1,-1) \rightarrow S_4=(1,1) 
\rightarrow S_1=(-1,1)$.
This can be represented by
a one-step Markov transition matrix $T$, where $T_{ij}$ element is 1
if state $i \rightarrow j$, and 0 otherwise. 
The transition matrix for R8 is:
\begin{equation}
\label{eq-R8-transition}
T=
\begin{array}{cc}
i \backslash j  & S_0 \hspace{0.1in} S_1 \hspace{0.1in} S_2 \hspace{0.1in} S_3 \\
\begin{array}{c}
S_0 \\
S_1 \\
S_2 \\
S_3
\end{array}
&
\begin{pmatrix}
0 & 0 & 1 & 0 \\
1 & 0 & 0 & 0 \\
0 & 0 & 0 & 1 \\
0 & 1 & 0 & 0 
\end{pmatrix}
\end{array}
\end{equation}
The matrix $T$ can be called ``row-wise one-hot"
because there is only one 1 (hot) in a row with the rest being 0s.
Eq.(\ref{eq-R8-transition}) is special because it is also
``column-wise one-hot", but other variants in Fig.\ref{fig3-R8}
do not have this property. When a transition matrix is one-hot
both row-wise and column-wise, it is a permutation matrix \citep{perm}.

The eigenvalues and eigenvectors of the transpose of a Markov 
transition matrix provide rich information of both limiting 
and transient dynamics. For example, the ``left" eigenvector
corresponding to the largest eigenvalue (Perron-Frobenius
eigenvector), after normalization, is the limiting stationary 
distribution (see, e.g., Appendix of \citep{wli-gene}). 
If a transition matrix is row-wise one hot,
but not column-wise, there is a zero eigenvalue, which is
an indication that one state is only a transient state towards
a limiting attractor. When all these zero-sum columns are
excluded, the rest of the matrix can be decomposed to disjoint
permutation subblocks. Each $p \times p$ submatrix contributes
$p$ eigenvalues of the form $e^{i 2\pi k/p}, i=0,1,\cdots,p-1$.

We use the following shorthand notations to indicate the limiting
dynamics: $F_1$ for one single global fixed-point
(and $F_2$ for two fixed-point attractors, and $F_3$, $F_4$
for three and four limiting fixed points); 2C for either
one or two limiting two-cycles; M (mixed) for existence
of both fixed-point(s) and two-cycle(s) attractor; 3C, 4C
for existence of three-, four-cycle.  Each limiting dynamical 
behavior has a unique set of eigenvalues from the transpose of 
the  Markov transition matrix,
and their relationship is summarized in Supplement
Material, item 4.

\subsection{Programs used}
\label{Msec:program}

\indent

Most computer runs of the MPN models were carried out
by {\sl R (https://www.r-project.org/}) programs written by the authors.
The UMAP (uniform manifold approximation and projection) run was
carried out by the $R$ UMAP package
({\sl https://cran.r-project.org/packages=umap}).
Our {\sl R} functions and scripts used in this paper is distributed
at {\sl https://github.com/wlicol/MPN2/}. 

\section{Results}

\indent

In a complex network with a large number of nodes and links,
if a subnetwork appears frequently or plays a more important role,
this subgraph may be called a prototype, pattern, a building block, a cluster,
a module, a motif, etc.  Two-node network motifs are commonly dismissed as too simple,
because there are only the possibilities of unidirectional and bidirectional links.
The transcriptional regulation networks usually
have transcription factor (master), which is a product from another gene,
acting on a to-be-regulated gene (slave) \citep{ocone}.

With self-loops, we can expand the above five classes of models to 39
(Fig.\ref{fig2} plus Fig.\ref{fig1}(B)  and Table \ref{table1}). 
There are two types of self-loop:
autocatalytic and self-regulating
(see Supplement Material, item 5). For unidirectional models
(commensalism and amensalism), there are six new models
(autocatalytic donor,
autocatalytic receiver, 
self-regulating donor, 
self-regulating receiver
autocatalytic donor and self-regulating receiver,
autocatalytic receiver and self-regulating donor,
two autocatalyses,
two self-regulation)
(see Supplement Material, item 5).
Combining them with the basic model, we expand the number of MPNs to 9.
The situation for predator-prey models is similar as there is an
asymmetry between the two nodes. One may just change the name donor
to predator, and name receiver to prey. The mutualism and competition
models are symmetric between the two nodes. Therefore, only five new models
are added. Overall, the number of MPN models with self-loop is
9+9+9+6+6=39.

\begin{table}[H] \footnotesize
\begin{center}
\begin{tabular}{c|c|c|c|c|c}
\hline
topology& with sign  & R & a b c d & dynamics(V2) & notes  \\
\hline
mutual &mutualism &53 &  0 1 1 0 & M   & Thomas' positive feedback, Hopfield  \\
regulating &&26 &  -1 1 1 0 & M &  one self-regulation \\
 &&25 &  -1 1 1 -1 &  M  &   two self-regulations \\
 &&54 &  0 1 1 1 & F$_2$   &  one autocatalysis \\
 &&27 &  -1 1 1 1 & F$_1$   &  one autocatalysis, one self-regulation \\
 &&81 &  1 1 1 1 & F$_2$   & two autocatalyses \\
\hline
 &competition&29 &  0 -1 -1 0 & M  & Thomas' positive feedback,Hopfield  \\
 &&2 &  -1 -1 -1 0 & M   &  one self-regulation \\
 &&1 &  -1 -1 -1 -1 & 2C  &  two self-regulations \\
 &&30 &  0 -1 -1 1 & F$_2$   & one autocatalysis  \\
 &&3 &  -1 -1 -1 1 & M    & one autocatalysis, one self-regulation \\
 &&57 &  1 -1 -1 1 & F$_3$   & two autocatalyses  \\
\hline
 &predator-prey&35 &  0 -1 1 0 & 4C  & Thomas' negative feedback \\
 &&8 &  -1 -1 1 0 & 3C  &   self-regulating prey  \\
 &&20 &  -1 1 -1 0 & 3C  &  self-regulating predator    \\
 &&7 &  -1 -1 1 -1 & 3C  &  two self-regulations  \\
 &&36 &  0 -1 1 1 & F$_1$  &  autocatalytic predator  \\
 &&48 &  0 1 -1 1 & F$_1$  &  autocatalytic prey  \\
 &&9 &  -1 -1 1 1 & 2C & self-regulating prey+ autocatalytic predator  \\
 &&21 &  -1 1 -1 1 & F$_1$  & autocatalytic prey + self-regulating predator \\
 &&63 &  1 -1 1 1 & F$_2$  & two autocatalyses  \\
\hline
master- &commensalism &44 &  0 0 1 0 & F$_1$  &   \\
slave &&23 &  -1 1 0 0 & F$_1$  & self-regulating commensal \\
 &&24 &  -1 1 0 1 & M  &  self-regulating commensal+ autocatalytic host \\
 &&51 &  0 1 0 1 & F$_2$  &  autocatalytic host \\
 &&17 &  -1 0 1 0 & 2C  &  self-regulating host \\
 &&16 &  -1 0 1 -1 & 2C  & two self-regulations  \\
 &&18 &  -1 0 1 1 & 2C  &  autocatalytic commensal+self-regulating host\\
 &&45 &  0 0 1 1 & F$_1$  &  autocatalytic commensal  \\
 &&72 &  1 0 1 1 & F$_3$  &  two autocatalyses \\
\hline
 &amensalism &32 &  0 -1 0 0 & F$_1$  &   \\
 &&5 &  -1 -1 0 0 & 2C  &  self-regulating victim \\
 &&6 &  -1 -1 0 1 & M   &  self-regulating victim + autocatalytic dominator \\
 &&33 &  0 -1 0 1 & F$_2$  &  autocatalytic dominator  \\
 &&4 &  -1 -1 0 -1 & 2C   &  two self-regulations \\
 &&11 &  -1 0 -1 0 & 2C  &  self-regulating dominator  \\
 &&12 &  -1 0 -1 1 & 2C  &  autocatalytic victim+ self-regulating dominator \\
 &&39 &  0 0 -1 1 & F$_2$  & autocatalytic victim  \\
 &&60 &  1 -1 0 1 & F$_3$  &  two autocatalyses \\
\hline
\end{tabular}
\end{center}
\caption{ \label{table1}
Each of the 39 MPN models is represented by one rule number
and the corresponding parameters $(a,b,c,d)$.
These rules are
first organized by whether it is bidirectional (mutual regulating)
or unidirectional (master-slave). Then the first group is split
into three more subclasses (mutualism, competition, predator-prey)
by the sign of the interaction; and the second group is split
into two subclasses (commensalism, amensalism). Lastly, by adding
self-loops, a total of 39 logical structures are listed.}
\end{table}

\large

\subsection{Different variants may exhibit different dynamics}
\label{Rsec:difvar}

\indent

Table \ref{table2-39rules} lists the limiting dynamics of the 39 MPN models 
(folded rule space) with  
6 variants, both synchronous and asynchronous updating (the last column
will be discussed in the Discussion section). For V1 variant,
39 independent rules can be reduced to 21 rules by the discrete gauge transformation
$Z_2$ (see Supplement Material, item 6). The representatives 
of these 21 rules are marked by asterisk in Table \ref{table2-39rules}.
Under the $Z_2$, any ``mutualism" model is equivalent to
a ``competition" model, and a ``commensalism" model equivalent to a ``amensalism"
model. Within the predator-prey models, three are self equivalent, and three
other models are not discrete-gauge-transformation-equivalent to three other models.
It leads to 9+6+3+3=21 independent variant-1  models.

Viewing from Table \ref{table2-39rules}, bipolar variants (e.g. V1, V2=V3) tend to have
a richer range of dynamical behaviors than binary variants (e.g. V4=V7, V6), which
have more fixed point limiting dynamics, as previously reported in \citep{huerta}. 
Fig.\ref{fig3-R8} illustrates this
by the R8 (competition with one self-regulation). The grey lines show the
threshold lines for updating $x$, and the pink lines those for updating $y$.
The states on the threshold lines may map differently in different variants.
It is clear that the threshold lines for bipolar variants are located
differently from those for binary variants. Therefore, it is understandable
that the limiting dynamics could be different
(Supplement Material, item 7, shows the threshold value
introduced by a transformation from bipolar to binary variant models).

\begin{table}[H] \footnotesize
\begin{center}
\begin{tabular}{c|c|ccc|cc|cc|ccc|ccc|c}
\hline
rule & links & \multicolumn{3}{c|}{equivalentR} &\multicolumn{4}{c|}{bipolar  variants} & \multicolumn{6}{c|}{ binary variants} & 3-level\\
\hline
R & a b c d & T12 &$Z_2$ & T12& \multicolumn{2}{c|}{syn} & \multicolumn{2}{c|}{asyn} & \multicolumn{3}{c}{syn} & \multicolumn{3}{c|}{asyn} &keep\\ 
\cline{6-15}
 & & & & +$Z_2$& V1 & V2 & V1A & V2A & V4 & V5 &  V6  & V4A & V5A & V6A & singular\\
 & & & & & & =V3 & & =V3A & =V7 & &  & & &  & state\\
\hline
1* & -1  -1 -1 -1 & 1 & 25 & 25 & M & 2C & F$_2$ & 3C  & F$_1$& 2C & F$_1$ & F$_1$ & 3C & F$_1$ & M=2C+F$_1$\\
2* & -1  -1 -1 0 & 28 & 26 & 52 & M & M  &F$_2$ & F$_1$ &F$_2$ &M  & F$_1$ & F$_2$ & F$_1$ & F$_1$ & 2C\\
3* & -1  -1 -1 1 & 55 & 27 & 79 & F$_2$&M & F$_2$ & M &F$_2$ &F$_1$  & F$_2$&  F$_2$ & F$_1$ & F$_2$  & four 2C \\
4* & -1 -1  0 -1 & 10 & 22 & 16 & 2C & 2C & 2C & 2C  &F$_1$ &2C & F$_1$ &F$_1$ &2C & F$_1$ & 2C\\
5* & -1 -1 0 0 & 37 & 23 & 43 & F$_2$&2C  &F$_2$ & 2C  &F$_2$ &F$_1$  & F$_1$ & F$_2$ &F$_1$  & F$_1$ & M=2C+F$_1$ \\
6* & -1 -1 0 1 & 64 & 24 & 70 & F$_2$&M & F$_2$&M &F$_2$ &F$_1$  & F$_2$ & F$_2$ &F$_1$  & F$_2$ & two 2C\\
7* & -1  -1 1 -1 & 19 & 19 & 7 & 4C &3C & 2C & 2C  &F$_1$ &3C  & F$_1$ & F$_1$ &2C  & F$_1$ & 8C\\
8* & -1  -1 1 0 & 46 & 20 & 34 & 4C & 3C & 2C & 2C  &F$_2$ &F$_1$ & F$_1$ & F$_2$ &F$_1$ & F$_1$ & two 3C \\
9* & -1 -1 1 1 & 73 & 21 & 61 & F$_2$&2C & F$_2$&2C  &F$_2$ &F$_1$ & F$_2$  & F$_2$ &F$_1$ & F$_2$ &  F$_1$\\
11* &  -1 0 -1 0 &31 & 17 & 49 & 2C& 2C & 2C & 2C & F$_2$& 2C & F$_1$ & F$_2$& 2C & F$_1$  & 2C\\
12* &  -1 0 -1 1 &58 &18 &76 & 2C& 2C & 2C & 2C & F$_2$& 2C & F$_2$ & F$_2$& 2C & F$_2$ & two 2C\\
16 &  -1 0 1 -1 &22 &10 &4 & 2C& 2C & 2C & 2C & F$_1$& 2C  & F$_1$ & F$_1$& 2C  & F$_1$  & 2C\\
17 &  -1 0 1 0 &49 &11 &31 & 2C& 2C & 2C & 2C & F$_2$& 2C  & F$_1$ & F$_2$& 2C  & F$_1$ & 2C\\
18 &  -1 0 1 1 &76 &12 &58 & 2C& 2C & 2C & 2C & F$_2$& 2C  & F$_2$ & F$_2$& 2C  & F$_2$  & two 2C\\
20 & -1  1 -1 0 &34 &8 &46 &4C& 3C & 2C & 2C & F$_1$& 3C  & F$_1$ & F$_1$& 2C  & F$_1$ & two 3C\\
21 & -1 1 -1 1 &61 &9 &73 & F$_2$& F$_1$ & F$_2$& F$_1$ &  F$_2$& F$_1$  &F$_1$ & F$_2$& F$_1$  &F$_1$  & F$_1$ \\
23 & -1 1 0 0 &43 &5 &37 & F$_2$& F$_1$ &F$_2$& F$_1$  & F$_2$& F$_1$ & F$_1$ & F$_2$& F$_1$ & F$_1$ & M=2C+F$_1$\\ 
24 & -1 1 0 1 &70 &6 &64 & F$_2$& M & F$_2$& M & F$_2$& F$_1$  & M & F$_2$& F$_1$  & M & two 2C\\
25 & -1  1 1 -1 &25 &1 &1 & M& M & F$_2$ & F$_1$ & M& M  & M & F$_2$ &  F$_1$ & F$_1$   & M=2C+F$_1$\\
26 & -1  1 1 0 &52 &2 &28 &M& M & F$_2$ & F$_1$ &F$_2$& F$_1$  &  M &  F$_2$ &  F$_1$ & F$_1$ & 2C\\
27 & -1 1 1 1 &79 &3 &55 & F$_2$& F$_1$ & F$_2$& F$_1$  & F$_2$& F$_1$ &  M & F$_2$& F$_1$ &  M & four 2C\\
29* & 0  -1 -1 0 &29 &53 &53 & M& M &F$_2$& F$_2$  &F$_3$& M  & F$_1$ & F$_3$ & F$_2$ & F$_1$ & M=2C+F$_2$\\
30* & 0 -1 -1 1 &56 &54 &80 & F$_2$ & F$_2$ &F$_2$& F$_2$  & F$_3$& F$_2$ & F$_2$ & F$_3$& F$_2$ & F$_2$ & F$_2$ \\
32* & 0 -1 0 0 &38 &50 &44 & F$_2$& F$_1$ &F$_2$& F$_1$  &F$_3$& F$_1$  & F$_1$  &F$_3$& F$_1$  & F$_1$  & F$_1$\\
33* & 0 -1 0 1 &65 &51 &71 & F$_2$& F$_2$ & F$_2$& F$_2$   &F$_3$& F$_1$  & F$_2$ & F$_3$& F$_1$  & F$_2$  & F$_2$\\
35* & 0  -1 1 0 &47 &47 &35 & 4C& 4C & 2C & 2C  &F$_2$& F$_1$ & F$_1$  & F$_2$& F$_1$ & F$_1$ & 4C\\
36* & 0 -1 1 1 &74 &48 &62 & F$_2$& F$_1$ & F$_2$& F$_1$  & F$_2$& F$_1$ & F$_2$ & F$_2$& F$_1$ & F$_2$ & 6C \\
39* & 0 0 -1 1 &59 & 45 & 77 &F$_4$ & F$_2$ & F$_4$ & F$_2$   & F$_4$ & F$_2$ & F$_2$ & F$_4$ & F$_2$ & F$_2$  &F$_3$ \\
44 & 0 0 1 0 & 50 & 38 & 32 &F$_2$ &F$_1$ &F$_2$ &F$_1$   &F$_3$ &F$_1$ & F$_1$ & F$_3$ &F$_1$ & F$_1$ & F$_3$\\
45 & 0 0 1 1 & 77 & 39& 59 &F$_4$ &F$_1$ &F$_4$ &F$_1$   &F$_3$ &F$_1$  &F$_2$ & F$_3$ &F$_1$  &F$_2$ & F$_3$\\ 
48 & 0 1 -1 1 & 62 &36 &74 & F$_2$& F$_1$ &F$_2$& F$_1$  &F$_3$& F$_2$ & F$_1$ & F$_3$& F$_2$ & F$_1$ & 6C\\
51 & 0 1 0 1 & 71 &33 &65 &F$_2$ &F$_2$ &F$_2$ &F$_2$   &F$_3$ &F$_1$ & F$_2$ & F$_3$ &F$_1$ & F$_2$ & F$_2$\\
53 & 0  1 1 0 & 53 &29 &29 &M & M &F$_2$ &F$_2$  &F$_2$ & F$_1$ & M & F$_2$ & F$_1$ & F$_2$  & F$_4$\\
54 & 0 1 1 1  &80 &30 &56 & F$_2$ & F$_2$ & F$_2$ & F$_2$  & F$_2$ & F$_1$ & F$_2$ & F$_2$ & F$_1$ & F$_2$  &F$_2$ \\
57* & 1 -1 -1 1 & 57 & 81 & 81 & F$_4$ & F$_3$ & F$_4$ & F$_3$  & F$_4$ & F$_3$  & F$_3$ & F$_4$ & F$_3$  & F$_3$ & F$_3$\\
60* & 1 -1 0 1 & 66 & 78 & 72 & F$_4$ & F$_3$ &F$_4$ & F$_3$  & F$_4$ & F$_2$ & F$_3$ & F$_4$ & F$_2$ & F$_3$ & F$_2$\\ 
63* & 1 -1 1 1 &75 &75 &63 &F$_4$ & F$_2$ &F$_4$ & F$_2$   & F$_3$ & F$_2$ & F$_2$ & F$_3$ & F$_2$ & F$_2$ & 8C\\ 
72 & 1 0 1 1 & 78 & 66 & 60 &F$_4$ & F$_3$ & F$_4$ & F$_3$  & F$_3$ & F$_1$ &F$_3$ & F$_3$ & F$_1$ &F$_3$ & F$_2$\\ 
81 & 1 1 1 1 & 81 & 57 & 57 & F$_4$ & F$_2$ &F$_4$ & F$_2$   & F$_2$ & F$_1$ & F$_2$ & F$_2$ &  F$_1$ & F$_2$ & F$_3$\\
\hline
\end{tabular}
\end{center}
\caption{ 
\label{table2-39rules}
Information on 39 MPN rules (excluding $N=1, N=0$ rules 
in Supplement Material, item 1).
R: rule number; $(a,b,c,d)$: four model parameters; T12: rule after switching two 
genes transformation; $Z_2$: after $b \rightarrow -b$, $c \rightarrow -c$ transformation
(discrete gauge transformation, see Supplement Material, item 6); 
T12+$Z_2$: after both T12 and $Z_2$ ;
V1,V2=V3,V4=V7, V5,V6: dynamical behavior under the seven variants of the model:
F$_4$, F$_3$, F$_2$, F$_1$ are fixed points with 4,3,2,1 states in the limiting set,
M is mixture of fixed-point and 2-cycle, 2C/3C/4C are 2/3/4-cycles. 
The 21 non-equivalent rules by T12, $Z_2$ and T12+$Z_2$ for V1 are marked by asterisk.
}
\end{table}
\large

\begin{figure}[H]
\begin{center}
  \begin{turn}{-90}
  \end{turn}
    \includegraphics[width=0.9\textwidth]{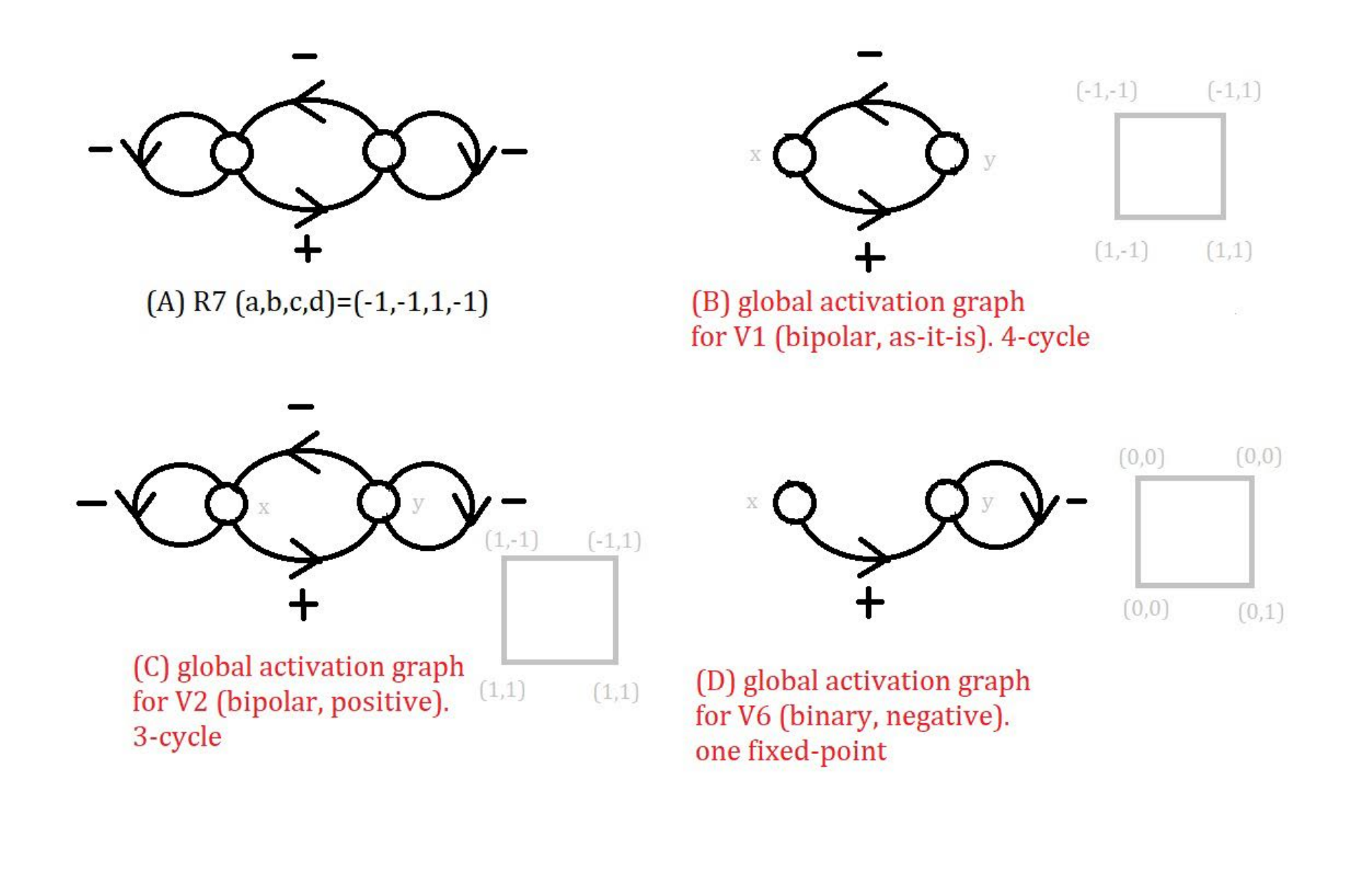}
\end{center}
\caption{
\label{fig4-GIG}
Illustration of the construction of global
interaction graph by the MPN model R7, for three different variants.
(A) The original regulatory graph where the sign of an arrow is determined
by the parameters $(a,b,c,d)$.
(B) The reconstructed global interaction graph for V1 (bipolar, as-it-is).
The four network states marked at the four corners of a square
represent four mappings:
$(1,-1)$ at lower-left for $(-1,-1) \rightarrow (1,-1)$,
$(1,1)$ at lower-right for $(1,-1) \rightarrow (1,1)$,
$(-1,-1)$ at upper-left for $(-1,1) \rightarrow (-1,-1)$,
and $(-1,1)$ at upper-right for $(1,1) \rightarrow (-1,1)$.
Moving from left to right, the first bits are unchanged and the
second bits increase. It implies $x$ has positive impact on $y$.
Moving from bottom to top, the second bits are unchanged whereas
the first bits decrease. It implies $y$ has a negative impact on $x$.
(C) similar to (B) for variant-2 (bipolar positive).
(D) similar to (B) for variant-6 (binary negative).}
\end{figure}

\subsection{On connection between signed regulatory structure and dynamics}
\label{Rsec:thomas}

\indent

Rene Thomas proposed two hypothesis relating the ``logical structure"
of a subnetwork with feedback loop and its dynamics \citep{thomas81}:
(1) the presence of at least one negative loop (odd number of
inhibitory links) is a necessary condition for a cyclic limiting dynamics;
(2) the presence of at least one positive loop (even number of
inhibitory links) is a necessary condition for multiple limiting
attractors (multiple steady states).

However, here we have more than one ``logical structures" for a MPN model.
Take R7 for example, it is a predator-prey model with two negative self loops
(see Fig.\ref{fig4-GIG}(A)). However, if we construct the global interaction
graph by the algorithm in \citep{richard}
(see Supplement Material, item 8), 
we have a predator-prey without
self loops for variant-1, commensalism with a self-regulating self loop
at commensal for variant-6, same graph as the original one for variant-2
(Fig.\ref{fig4-GIG}(B),(D), (C)).

If we check Thomas' hypotheses on the global interaction graph
(see Supplement Material, item 8),
there is no counter examples. R7/V1 has a limiting 4-cycle,
R7/V2 has a limiting 3-cycle, both contain negative loops.
R7/V6 does not have cyclic dynamics and does not have multiple
attractors, so it is not covered by Thomas' hypothesis.

To use Thomas' hypothesis to predict the limiting dynamics from
a given MPN model has several shortcomings. First, the existence
of  positive/negative loops is not a sufficient condition.
Second, the construction of global interaction graph
in Fig.\ref{fig4-GIG}(B,C,D) requires a one-step running
of the model to get the network state transition table.
As we have seen from Sec.~\ref{Msec:spectrum}, if the state
transition table is available, we should be able to predict the
limiting dynamics by the set of eigenvalues. In some sense, we have
already run a simulation (for one step), and it is no longer
a theoretical prediction. As spectral approach already
provides a complete solution, global activation graphs,
on the other hand, have only much limited success. 
Finally, it has already been pointed out
that there is a violation of Thomas' hypothesis in synchronous
updating \citep{richard}.

In the remaining of this subsection, we move away from Thomas'
hypotheses by not using the global activation graph, but ask this 
question: are there some trends in the relationship between
the logical structure as defined by the MPN model 
(e.g., Fig.\ref{fig1}(B), Fig.\ref{fig2}, Fig.\ref{fig4-GIG}(A)),
and the limiting dynamics (at least in some variants)? 

\begin{table}[H] \normalsize
\begin{center}
\begin{tabular}{c|c|cc|c|c}
\hline
rule & a b c d &  $Z_2$& T12+$Z_2$ & V1 dyn & description \\
\hline
39 & 0 0 -1 1 &   45 & 77 &F$_4$ &   amensalism+vic\_autoC \\
60 & 1 -1 0 1 &  78 & 72 & F$_4$ &  two autoC \\ 
57 & 1 -1 -1 1 &   81 & 81 & F$_4$ &  two autoC \\
63 & 1 -1 1 1 &  75 &63 &F$_4$ &  two autoC\\ 
\multicolumn{2}{l}{45} &  &  &F$_4$ &  commensalism+com\_autoC \\
\multicolumn{2}{l}{72,81} &  &  &F$_4$ &   two autoC\\
\hline
3 & -1 -1 -1 1 &   27 & 79 & F$_2$ & competition+1autoC  \\
30 & 0 -1 -1 1 & 54 &80 & F$_2$ & competition+1autoC \\
9 & -1 -1 1 1 &  21 & 61 & F$_2$  & predator-prey+1autoC \\
36 & 0 -1 1 1 &  48 &62 & F$_2$ &   predator-prey+1autoC \\
5 & -1 -1 0 0 & 23 & 43 & F$_2$  & amensalism+no\_dom\_self\_reg \\
6 & -1 -1 0 1 &  24 & 70 & F$_2$  & amensalism+no\_dom\_self\_reg\\
32 & 0 -1 0 0 & 50 &44 & F$_2$ &  amensalism+no\_dom\_self\_reg \\
33 & 0 -1 0 1 & 51 &71 & F$_2$ & amensalism+no\_dom\_self\_reg \\
\multicolumn{2}{l}{23,24,51}&   & &F$_2$ & commensalism+no\_host\_self\_reg \\
\multicolumn{2}{l}{21,48}&    & & F$_2$ & predator-prey+autoC\\
\multicolumn{2}{l}{27,54}&    & & F$_2$ & mutualism+autoC\\
\multicolumn{2}{l}{44}&   & & F$_2$ &  commensalism+no\_host\_self\_reg \\
\hline
1 & -1  -1 -1 -1 &    25 & 25 & M & competition+no\_autoC \\ 
2 & -1  -1 -1 0 &  26 & 52 & M  & competition+no\_autoC \\
29 & 0  -1 -1 0 &  53 &53 & M &competition+no\_autoC \\
\multicolumn{2}{l}{25,26,53} &  & & M   &mutualism+no\_autoC\\
\hline
4 & -1 -1  0 -1 &   22 & 16 & 2C &   amensalism+dom\_self\_reg \\
11 &  -1 0 -1 0 &  17 & 49 & 2C  &   amensalism+dom\_self\_reg \\
12 &  -1 0 -1 1 &  18 &76 & 2C   &  amensalism+dom\_self\_reg \\
\multicolumn{2}{l}{18 }&   & &  2C & commensalism+host\_self\_reg \\
\multicolumn{2}{l}{16,17 } &  & &2C &   commensalism+host\_self\_reg\\
\hline
7 & -1  -1 1 -1 &   19 & 7 & 4C  &   predator-prey+no\_autoC  \\
8 & -1   -1 1 0 &   20 & 34 & 4C  & predator-prey+no\_autoC \\
35 & 0  -1 1 0 &   47 &35 & 4C&  predator-prey+no\_autoC \\
\multicolumn{2}{l}{20 } &  & & 4C &  predator-prey+no\_autoC \\
\hline
\end{tabular}
\end{center}
\caption{ \label{table3-21V1}
The list of 21 non-equivalent models under T12, $Z_2$, T12+$Z_2$ for
V1 variant  organized by their V1 limiting dynamics,
in the F$_4$, F$_2$, M, 2C, 4C order;
i.e., models with the most number of fixed-points at the top,
and models with the longest cycle length at the bottom.
The description column is in term of the jargon used
in Table \ref{table1}.
}
\end{table}
\large

The fewer number of independent V1 models (21 models) provides us with 
the chance to address this question.
When the V1 models are grouped by their limiting dynamics,
with models with the largest number of fixed points at the top,
and models with the longest cycle length at the bottom in Table \ref{table3-21V1}, 
the original model graph can be examined.
We have observed these patterns:

\begin{itemize}
\item
Being predator-prey models without an autocatalysis is 
the sufficient and necessary condition for 4-cycle limiting dynamics.

\item
Having two autocatalyses is the sufficient, but not necessary,
condition for four fixed-points attractors (F$_4$).

\item
Being unidirectional (amensalism or commensalism) models with self-regulation 
on donor node is the sufficient and necessary condition for 2-cycle limiting
dynamics.

\item
Being competition or mutualism models without an autocatalysis 
is the sufficient and necessary condition for a mixed dynamics (2-cycles and fixed points).

\end{itemize}

The situation not covered above is F$_2$ (two fixed-points).
We can summarize the observed patterns in another way: (1) lacking of autocatalysis 
lead to longer cycles, whereas more (e.g. two) autocatalysese lead 
to more fixed points; (2) for unidirectional models, self-regulation in the
donor node tend to lead to cycles, whereas autocatalysis
in the receiving node tend to lead to more fixed points.

If we rearrange V2 models in Table \ref{table1} with F$_3$ dynamics 
near the top and 3C/4C dynamics near bottom, we find similar trend, 
i.e., more rules near the top have two autocatalyses, while more 
self-regulations are located at the bottom (results not shown).

\subsection{On the robustness of V1 models on point mutations}
\label{Rsec:rulespace}

\indent

\begin{figure}[H]
\begin{center}
  \begin{turn}{-90}
  \end{turn}
   \includegraphics[width=1.0\textwidth]{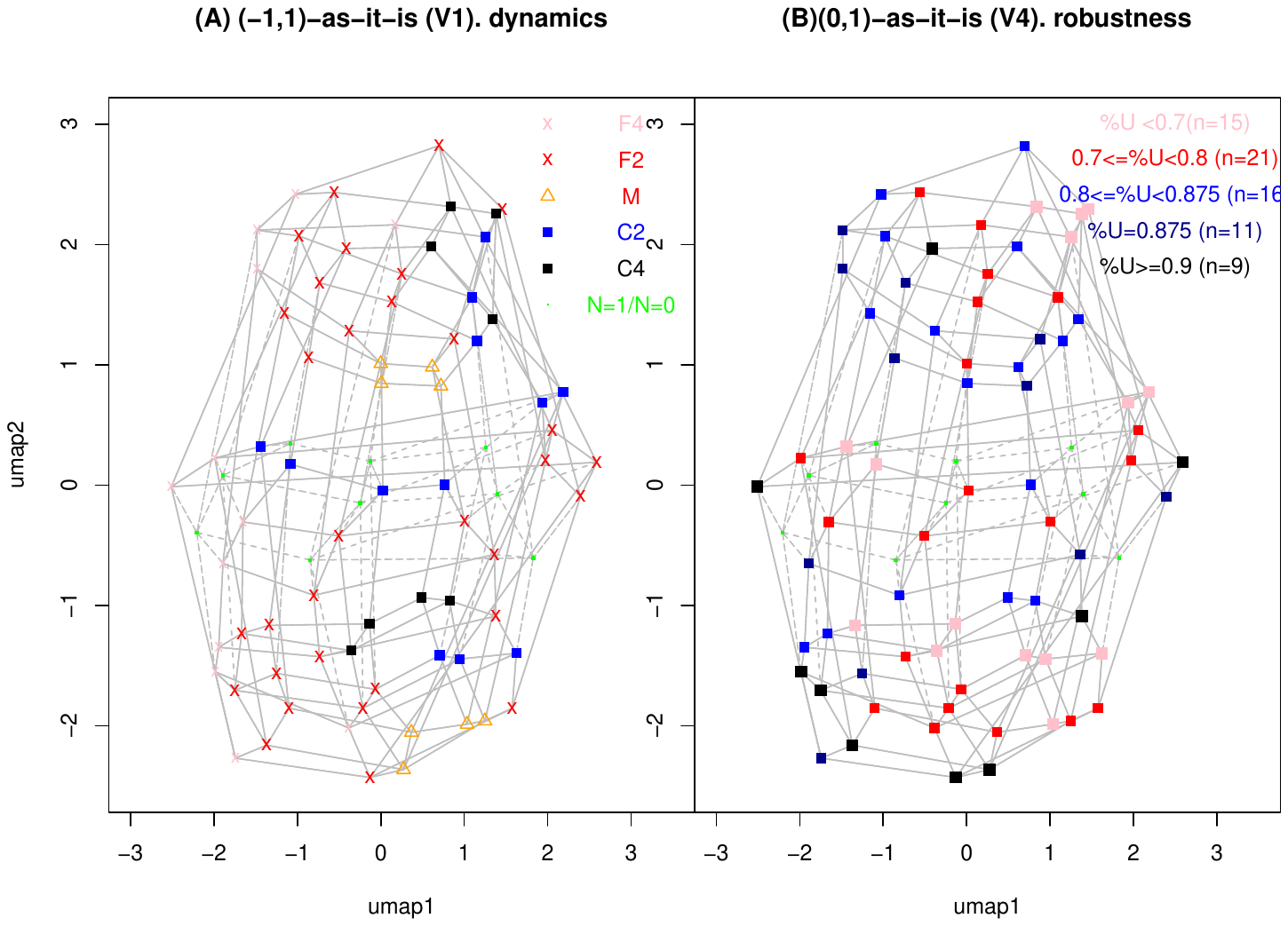}
\end{center}
\caption{
\label{fig5-UMAP}
UMAP of unfolded rule space of 81 two-node MPN models. 
The two axes in UMAP do not have a direct interpretable meaning.
When two rules, represented by two points, are linked, the Hamming distance between
the two is one. 
(A) nodes are colored by pink (V1 F$_4$ rules), red (V1 F$_2$ rules),
orange (V1 M rules), blue (V1 2-cycle rules), and black (V1 4-cycle rules).
The $N=1$ or $N=0$ rules are shown as green dots and link (one Hamming distance)
to them are in dashed lines.
(B) nodes are colored by the robustness (percentage of point mutation in rule 
that does not change the limiting state) in V4, with darker colors for
more robust rules. In legend of (B), \%U denotes the percentage of unchanged limiting dynamics.
}
\end{figure}

We study the MPN model space in its ``unfolded" version,
ignoring the equivalence between rules by node switching and by the discrete gauge
transformation (for V1 only). Folded space \citep{wli-ca} where these equivalences
are considered is, generally speaking, harder to visualize.
Fig.\ref{fig5-UMAP}(A) shows a projection of the MPN model space from 4-dimensional
(parameters $a,b,c,d$) grid to 2-dimension by UMAP (uniform manifold 
approximation and projection) \citep{umap}.

In Fig.\ref{fig5-UMAP}(A), fixed point rules (F$_4$ and F$_2$)
tend to be close to each other in the rule space, whereas high cycle rules (C2, C4,
and partially M) are closer to each other. This can be confirmed quantitatively.
Each rule can make a transition to another rule by a single point mutation in one
of the parameters of $(a,b,c,d)$. If (e.g.) $a=c=d=0, b \ne 0$, there are 7 nearest neighboring
rules in the rule space; if (e.g.) $a=b=c=d=1$, there are only 4 nearest neighbors.
Some mutations change to another rule in the same dynamic class, whereas other
mutations may change the dynamic behavior.

\begin{table}[H] \normalsize
\begin{center}
(A) \\
\begin{tabular}{c|ccccc|c}
\hline
from$\backslash$to & F$_4$ & F$_2$ & M & 2C & 4C & sum \\
\hline
F$_4$ &   24 (54.5\%)  & 16 & 0 & 4 & 0 & 44    \\
F$_2$ &   16 & 40 (47.6\%) & 8 & 12 & 8 & 84    \\
M     &   0  & 8 & 8 (40\%) & 4 & 0  & 20    \\
2C    &   4  & 12 & 4 & 24 (50\%)& 4  & 48    \\
4C    &   0  & 8 & 0 & 4 & 8  (40\%) & 20  \\
\hline
\end{tabular}

(B) \\
\begin{tabular}{c|ccc|c}
\hline
from$\backslash$to &  F & 2C+M & 4C & sum  \\
\hline
F       &  96 (75\%) & 24 & 8 & 128\\
2C+M    &  24 & 40 (58.8\%) & 4 & 68 \\ 
4C      &  8  & 4 & 8 (40\%) & 20   \\
\hline
\end{tabular}
\end{center}
\caption{ \label{table4}
(A) The number of point mutations (Hamming distance of 1) that change or do not 
change the V1 limiting dynamics (classified in five groups: F$_4$, F$_2$, M, 2C, 4C). 
The transition from/to a $N=1$ or $N=0$ rule is counted separately. 
(B) Similar to (A) but F$_4$ and F$_2$ are grouped as one group, M and 2C 
grouped as another group.
}
\end{table}
\large

Table \ref{table4}(A) shows the number of intra and inter class mutations
with five groups of distinct dynamical behaviors being considered:
F$_4$, F$_2$, M, 2C, 4C ( 
$N=1$ (i.e, $b=c=0$) or $N=0$ (i.e, $a=b=c=d=0$) rules are also counted in the table).
We found that 54.5\% of the non-attracting fixed-points rules (F$_4$)
mutate to another rule in F$_4$, 47.6\% of attracting fixed-point rules (F$_2$)
mutate to F$_2$, 40\% of fixed-point and 2-cycle mixture rules (M) mutations
are intra-class, 50\% 2-cycle rule mutations (and 40\% 4-cycle rule mutations) 
are intra-class.  Overall, 76 out of 168 mutations do not change the dynamic
behavior if we classify the rules in five groups.

The five groups of dynamical behavior can also be coarse-grained into three groups:
fixed points (combining F$_4$ and F$_2$), 2-cycle (combining M and 2C),
and 4-cycle remains as it was. The number of inter- and intra-class mutations
is shown in Table \ref{table4}(B). 
For fixed-point models, 
96 out of 128 mutations lead to another fixed-point model (75\%).
Similarly, for 2-cycle models (M and 2C), 
40 out of 68 mutations (58.8\%) are intra-class, and for high-cycle
models, 8 out of 20 mutations (40\%) are intra-class.

Borrowing a concept from complex dynamical systems, here we define a fixed-point 
MPN rule that is one Hamming distance 
(i.e., one of the parameters $a,b,c,d$
changes by one unit) away from a 4-cycle  rule 
to be an ``edge of high cycles" rule, mimicking the term ``edge of chaos" in
\citep{packard,langton}.
Of the 44 fixed-point rules, 
16 of them are edge-of-high-cycle by the above definition. In our 21 representative rules,
12 of them are fixed points rules (Table \ref{table3-21V1}). Of these 12, 4 are
edge-of-high-cycle 
rules (Rules 5, 9, 32, and 36). There are two ways a fixed point
model can mutate to a 4-cycle model: by transforming a unidirectional model to a bidirectional
model (e.g., R5 to R8, R32 to R35), or by removing an autocatalytic self-link 
(R9 to R8, R36 to R35). Since the models in Table \ref{table3-21V1} are sorted from 
highest number of fixed points near the top, to fewer number of fixed-points,
to mixing fixed-points and 2-cycle, to 2-cycles, and eventually to 4-cycles 
near the bottom, it is not surprising that these edge-of-high-cycle rules are 
indeed located in the middle, as boundary rules between top and bottom.

\subsection{Model robustness and neighborhood stability 
in V4 variant rule space}

\label{Rsec:rulespace2}

\indent

In the last section, we describe how the overall dynamics of a model
may change with a mutation in the rule parameter.
One can also ask for a given initial state,
how the limiting attractor starting from that state may change.
The V4 (binary as it is) variant, 
which is also equivalent to our ``binary difference rule" (V7) 
described in Supplement Material, item 3
is perfect for this purpose, as all rules but one (R25) are fixed-point rules
(see Table \ref{table2-39rules}).
If network rule is considered as ``genotype" and the limiting state as
gene-expression ``phenotype" (all other extra factors including 
``upstream gene" or ``environment" are included in the initial 
condition \citep{wagner94,wagner96}), this robustness is about resilience of 
gene-expression phenotype against modification of genotype with all 
other factors fixed (including the initial state). 

For each one of the four state values for $(x,y)$ for a given rule, 
there are 4-8 number of neighboring rules, resulting in 16-32 chances
to check robustness.
Rule 25, the only rule with possible 2-cycle limiting dynamics in V4,
has the lowest level of robustness:
only 37.5\% of the rule mutations 
do not change the limiting state. The next rule with the lower robustness
is R16/R22 (50\% rule change do not affect the limiting state).
The rules with the highest level of robustness (15 out of 16 rule mutations
do not change the limiting state) are R9/R73, R72/R78. 

Fig.\ref{fig5-UMAP}(B) shows the rule space where each rule is colored
by its level of robustness: 15 rules have less than 69\% of rule mutation
that do not change the limiting state (in pink), 21 rules with 70-79\% level
of robustness (in red), 16 rules with 80-85\% (in blue), 11 rules with 87.5\% of
the mutations that do not change the limiting state (in darkblue), and 9 superstable
rules (besides R9/R73 and R72/R78, R51/R71, R54/R80, R53) (in black).
When Fig.\ref{fig5-UMAP}(A) and Fig.\ref{fig5-UMAP}(B) are compared side by side
(note a subtle point that (A) is for V1 and (B) for V4),
we see a trend of complementary of colors, light color dots in Fig.\ref{fig5-UMAP}(A)
(fixed points) tend to be dark color in Fig.\ref{fig5-UMAP}(B) (more robust),
and vice versa. 

\begin{table}[H] \normalsize
\begin{center}
\begin{tabular}{c|ccc|cc}
\hline
dynamics & \multicolumn{5}{c}{\%unchanged limiting state w/ rule change(in V4)}\\
\cline{2-6} 
 class(in V1) &  $\le$ 0.7 & 0.71-0.76 & 0.79-0.82 & 0.821-0.88 & $\ge$ 0.9 \\
\hline
total & 17 & 18 &  20 &  14 & 12 \\
\hline
fixed-pt & 4 & 11 & 12 & 10 & 11 \\
 & \multicolumn{3}{c|}{(27)} & \multicolumn{2}{c}{ (21)}\\
\hline
2C or M & 9 & 7 & 4 & 4 & 1\\
4-cycle & 4 & 0 & 4 & 0 & 0\\
 & \multicolumn{3}{c|}{(28)} & \multicolumn{2}{c}{ (5)}\\
\hline
\end{tabular}
\end{center}
\caption{ \label{table5}
Count of rules by their robustness of limiting state (phenotype)
against point mutation in rule (genotype), measured by the percentage
of point mutations (for all possible initial states) that do not 
change the limiting state. Five groups are constructed  by this percentage 
to have a comparable number of rules per group. Each row is
a different limiting dynamics type. The four quadrants are a particular
partition of the limiting dynamics types (fixed-points vs. non-fixed-point),
and robustness (percentage equal or higher than 0.821 vs. lower).
The Fisher's test $p$-value for the 2-by-2 table is 0.00797.
}
\end{table}
\large

This hypothesis can be tested by checking a rule's
dynamics under variant V1 and its robustness under variant V4/V7
in Table \ref{table5}. If we group the counts
into a 2-by-2 table, V1 dynamics being fixed-point or not-fixed-point,
V3 robustness less than 0.821 or above 0.821, the odds-ratio is 9 (95\% confidence interval:
1.89-42.78), and Fisher test $p$-value is 0.00797. These provide 
statistical evidence that  cyclic dynamics in ``bipolar as it is"
variant (V1) is associated with a lower level of robustness in ``binary as it is"
variant (V4).

Previously, we examined the impact of rule change
(change of logical structure of a network model by adding/removing a link)
on dynamical behaviors (for V1 variant), and on the limiting state (for V4/V7
variant).  Here we examine a third type of robustness: the impact of initial state 
on the limiting state in V4, when rule is fixed.
If the limiting state is ``attracting", all initial states lead to
the same fixed point, and there is a robustness with respect to
initial state. On the other hand, if the limiting states are not
attracting, changing initial state will change the limiting state.

\begin{figure}[H]
\begin{center}
  \begin{turn}{-90}
  \end{turn}
   \includegraphics[width=0.9\textwidth]{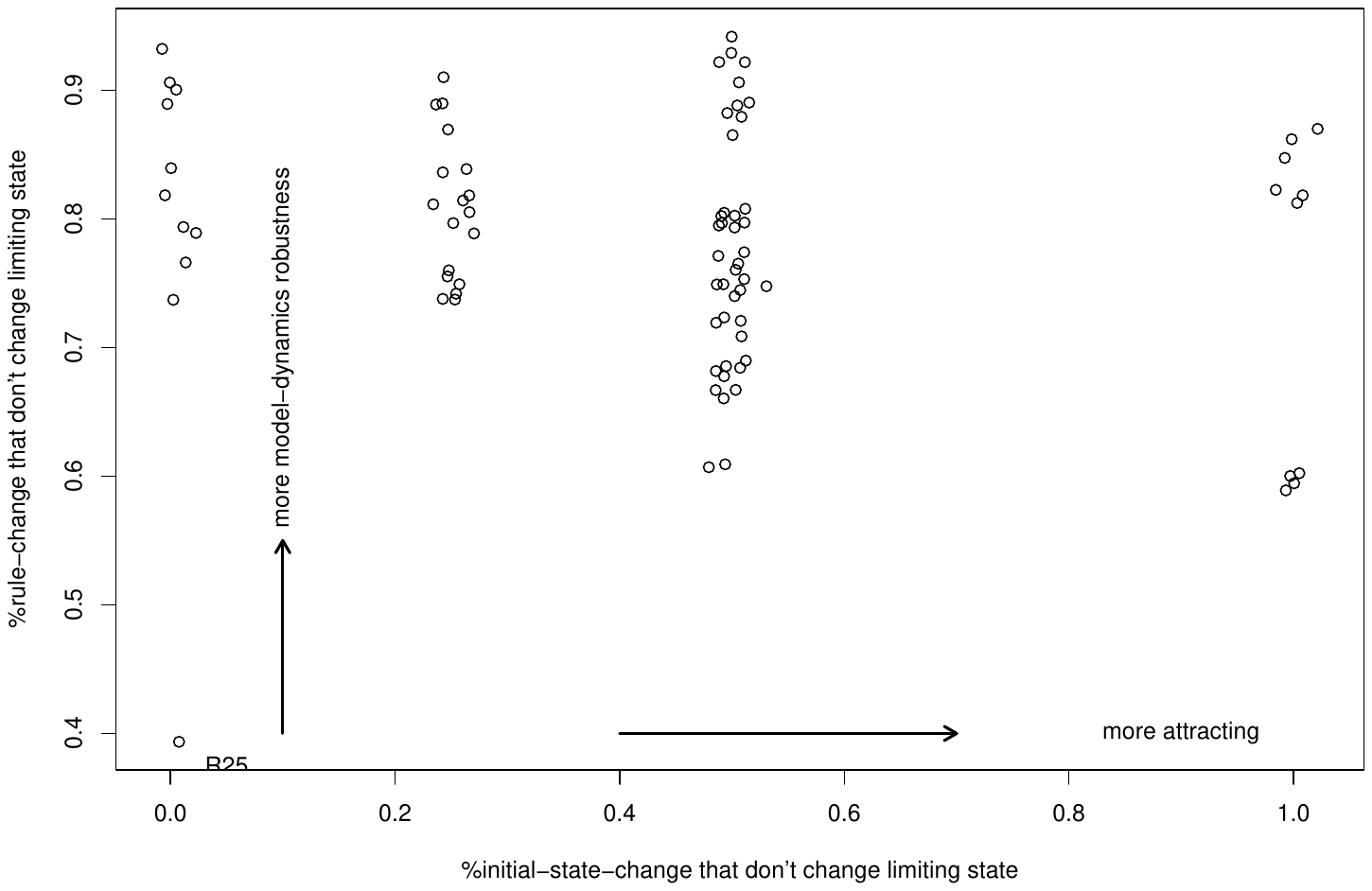}
\end{center}
\caption{
\label{fig6-robust}
Scatter plot of ($y$-axis) the percentage of point mutations in rules that does not change
the limiting state (V4) vs.  ($x$-axis) percentage of point mutation in initial state
that does not change the limiting state (V4). The $x$-axis can be understood as
a measure of robustness of phenotype against environmental factors, whereas the
$y$-axis as that against genotype changes.  There are total 72 points/rules in the plot
as $N=1$ or $N=0$ rules are excluded.
}
\end{figure}

There are only 4 initial states $(x,y)$ = (0,0), (0,1), (1,0), and (1,1).
Four Hamming-distance-1 changes on initial condition are carried out
and percentage of times when the limiting state is unchanged is calculated. 
With this percentage (for V4) as $x$-axis, and percentage of times rule change 
doesn't change the limiting set  (for V4) as $y$-axis, Fig.\ref{fig6-robust}
shows a weak negative correlation between the two. 
The correlation is not significant ($p$-value= 0.13 for Pearson correlation, 
and $p$-value =0.06 for Spearman correlation). 
However, once the lower-left outlier (R25) is removed,
the correlation $p$-values improve to 0.008 (Pearson) and 0.024 (Spearman).

\begin{figure}[H]
\begin{center}
  \begin{turn}{-90}
  \end{turn}
   \includegraphics[width=0.8\textwidth]{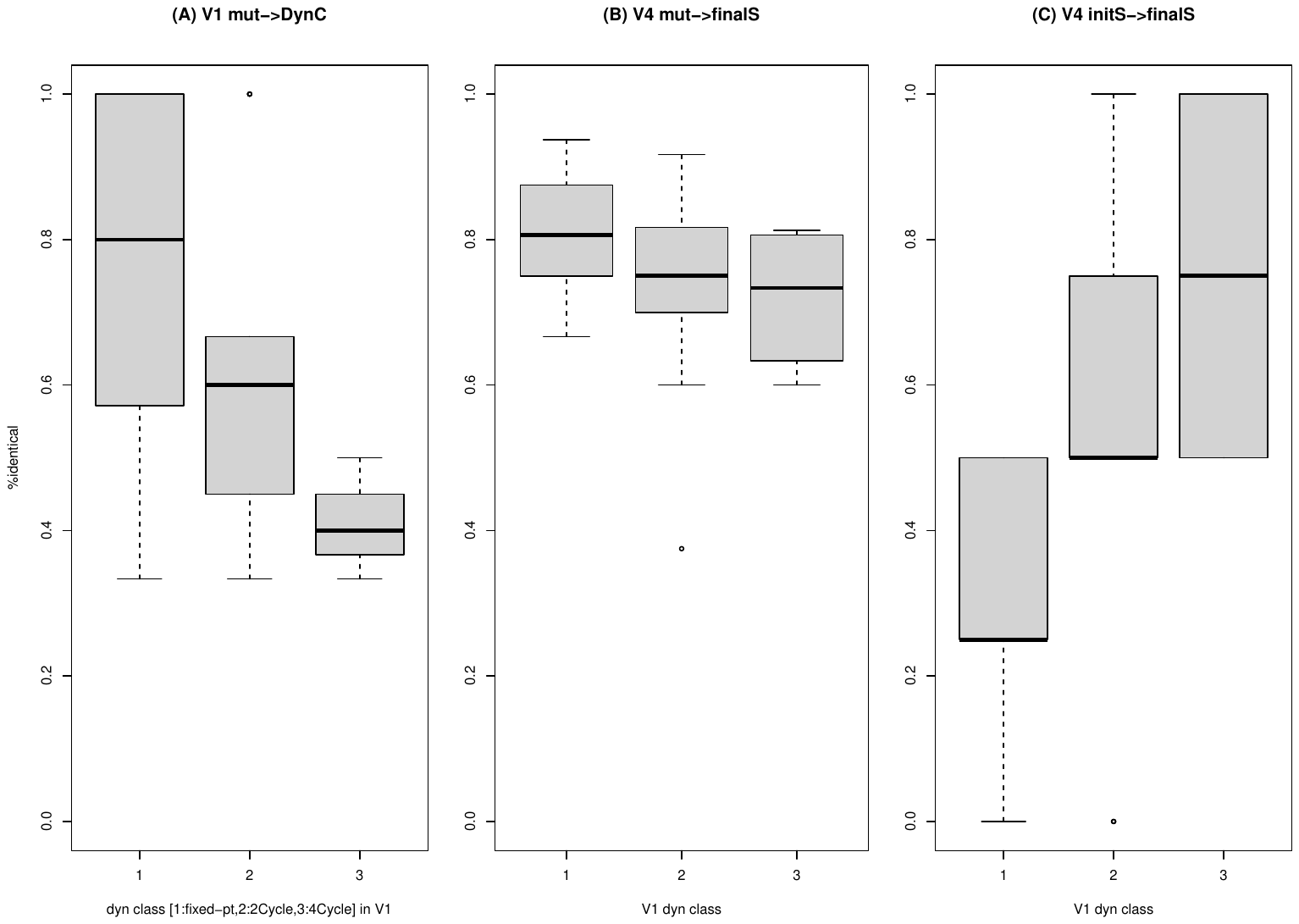}
\end{center}
\caption{
\label{fig7-boxplot}
Illustration of three different types of robustness of MPNs. 
In the $x$-axis, MPNs are grouped to three classes according to the dynamical 
behavior in V1: fixed points (F$_4$ and F$_2$), 2-cycles (2C and M), and 4-cycles (4C).
(A) Percentage of mutations in the rule that does not change the dynamic class in V1;
(B) Percentage of mutations in the rule that does not change the final states in V4;
(C) Percentage of changes in the initial state that does not change the final state in V4.
The percentages are calculated from 45 rules including 39 $N=2$ rules in 
Table \ref{table2-39rules}, as well as six $N=0$ or $N=1$ rules.
}
\end{figure}

The three types of robustness can be illustrated by the boxplots in 
Fig.\ref{fig7-boxplot}.
In these boxplots, both the 39 $N=2$ rules in  Table \ref{table2-39rules} and 
six $N=0$, $N=1$ rules, as discussed in Supplement Material, item 1, 
are used, and rules are grouped into three classes by
their limiting dynamics in variant-1 (V1): fixed points, two cycles (2C and M), and 4-cycles.
In Fig.\ref{fig7-boxplot}(A), the percentage of mutations in rules that do not change
the limiting dynamical behavior (in V1)  is shown in the $y$-axis.  In Fig.\ref{fig7-boxplot}(B), 
the percentage of mutations in rules that do not change the limiting state 
in variant-4 (V4) is shown in the $y$-axis. In both cases, fixed point class
is more robust with respect to rule changes.
Figure \ref{fig7-boxplot}(C) shows the percentage of initial states that do not
change the limiting state in variant-4. The trend is the opposite to those in
Fig.\ref{fig7-boxplot}(A,B): 
fixed-point (in V1) rules are less stable with respect
to perturbation in the initial conditions to its final state (in V4). 

\section{Discussion}

\indent

Despite the minimalistic approach of this work (model space of all two-node/gene
McCulloch-Pitts networks), we believe that it addresses extensively the
domain of 2-node networks, an attempt not undertaken before, to the best of our knowledge.
Some studies seem to address threshold models and ``Boolean cellular automata"
\citep{kurten1,kurten2}, but the emphasis was still not on all rules, but only 
on randomly selected ones.  Another work essentially classified the equivalence
of limiting dynamics instead of classifying the rules themselves \citep{glass75}.
One might expect that our choice of only two nodes would render the results 
trivial. Admittedly, some interesting phenomena are missing, 
for example, feed-forward loop regulation requires at least three node/genes 
\citep{mangan}.  However, we wish to show that these ``minimal complex systems"
may not be as trivial as one might first think due to the extreme 
nonlinearity and self feedback loops.

{\bf Relations to Hopfield, Kauffman, Wagner networks:}
Besides the MPNs, and along the same lines, we noticed
a number of other models that could be related:
(1) In Hopfield models \citep{hopfield82} 
the link strength from node-1 to node-2 is equal to that from node-2 to node-1,
and there is no self-link. Two of the 39 two-node MPNs are 
Hopfield models. These two Hopfield models are actually the
well known mutualism (both links are +1) and competition (both
links are $-1$) models. 

(2) Kauffman's random networks \citep{kauffman69,kauffman71} 
(Kauffman being greatly influenced by McCulloch himself \citep{kauffman})
use logic gates (e.g. AND, OR, XOR, etc.; for all 16 possible
logic gates for two inputs, see Supplement Material, item 9) 
to model the interaction between
nodes.  Logical gates are not easily parameterized as MP functions are, 
which makes a study of model space more difficult.  In the notation of 
$NK$ Kauffman model
\citep{kauffman-levin,weinberger,kauffman-book}, our work would match $N=2$, 
but only seven out the 39 MPN models are $K=1$, and ten are $K=2$,
with the rest having different number of incoming links for the two nodes.
Although MPNs can't model XOR (exclusive OR) gate, it is suggested
that XOR does not have a biological meaning \citep{raeymaekers}.

(3) Wagner models  \citep{huerta,pinho,wang-yf} (also see \citep{mjolsness}) 
are a population of MPNs where node links are randomly chosen 
\citep{wagner96,wagner-book05}, not restricted to $-1$,0,1 values. 
In Wagner models, the parameters in a  MPN are interpreted as ``genotype",
and the limiting dynamics as ``phenotype", and initial state of 
nodes are treated as an extra factor that may be upstream genes 
or other environment factors  \citep{wagner94,wagner96}.
Almost all previous studies on Wagner model were based on computer simulation
where a population of MP models are selected repeatedly by certain 
evolutionary criteria \citep{siegal,shin}.

{\bf Logic gates:}
As $x^{t+1}$ and $y^{t+1}$ are binary function of
$x^t$ and $y^t$ that are binary variables themselves, all MPNs can be
represented by logic gates (Supplement Material, item 9). Interestingly,
all N=2 MPNs for variant V1 are not really two-input gates, but
single-input gates (see Supplement Material, item 10 for details).
This is a situation of ``canalizing" \citep{kauffman-sa}. However,
for other variants (V2-V6), many MPNs are true two-input gates,
co-existing with the one-input gates (see Supplement Material,
item 11).

{\bf Recurrent neural networks (RNN):} 
Recurrent neural networks were proposed
to handle sequential data by using feedback loops to remember the past information
\citep{hinton92}. In the one-to-many version of RNN \citep{amidi}, one input
X updates the hidden variable H which both sends a self-loop to itself,
and updates an output Y. That Y replaces X as the next round input, and
a sequence of Y's are produced. This set up is essentially  a two-node
networks with one node H for hidden variable, and another node X/Y for both input
and output (which becomes input in the next step). We notice however that
(1) the updating of H and X/Y are not synchronous; (2) the weight of link
can be adjusted/learned by comparing the predicted and expected output Y.

{\bf Asynchronous updating:}
It has been suggested that asynchronous updating is more realistic
for describing gene regulations \citep{thomas91}. If we update $x$ first,
we may split one time step into two halves, first, $x^{t+0.5}=f(ax^t+by^t)$,
then $y^{t+1}= f(cx^{t+0.5}+d y^t)$, and finally reset $x^{t+1}=x^{t+0.5}$.
Each variant has its own asynchronous version, and call these V1A, V2A, $\cdots$, etc.
Updating $y$ first does not seem to make any difference.
Asynchronous updating never changes two nodes' value
at the same time, thus there is never a state transition across
the diagonal (or across off diagonal) as in the cases in Figs.
\ref{fig3-R8}
\citep{richard}.  
The only state transition allowed in asynchronous updating is along
the edge of the square in Figs. \ref{fig3-R8}
(or hypercube if the number of nodes is larger than 2).

The limiting dynamics of asynchronous updating of all variants
are shown in Table \ref{table2-39rules}.
Comparing the synchronous and asynchronous version of one variant,
for V1, there are 10 rules (out of 39 -- 25.6\%) that have different
dynamical group labels. In six of these rules, the M dynamics in
synchronous version becomes F$_2$ in asynchronous version; and for 
four rules, 4C dynamics becomes 2C dynamics. For V2, there are also
10 rules that exhibit slightly different dynamics between synchronous
and asynchronous versions, five in changing cycle length, and
five in changing from a mixed dynamics to purely fixed point dynamics.

For V4, V5, V6, there are even less number of rules that change dynamical
behavior when going from synchronous to asynchronous updating: only 1 (2.6\%), 6 (15.4\%), 4 (10.3\%)
rules change labels, respectively. Most of these changes are to lose the 2-cycle, and
change cycle length. In general, the change from synchronous to asynchronous
updating in our simplest MPN is less dramatic than expected (see \citep{assmann,saa} for 
discussion on other asynchronous networks). 

{\bf Sigmoid activation function:} As pointed out in \citep{snoussi93},
mapping value for points on the threshold line (e.g.
$ax+by=0$ or $cx+dy=0$) is special. One is often forced to make a decision,
which is one essential cause of different variants in this work (another
cause is the bipolar-binary difference). If we use a continuous function,
e.g. sigmoid/logistic function, $x^{t+1}= 2/(1+ exp(-ax^t-by^t))-1$
for bipolar models, and $x^{t+1}= 1/(1+ exp(-ax^t-by^t))$ for binary
models (that for $y^{t+1}$ is similar), then the node state values
could potentially be continuous. With this, there is no need
to distinguish V1, V2, V3 bipolar models (and V4, V5, V6 binary models).

However, a new problem is created: bipolar model may reach the 
``singular" state of 0, and binary model reach the ``singular" state of 0.5. 
This would potentially lead to three states (e.g., $[-1, 0, 1]$). 
The last column of Table \ref{table2-39rules} includes the limiting dynamics 
under sigmoid 
function.  Several models whose dynamics are all fix-points in V1-V6, nevertheless,
show cyclic dynamics (R36, R48, R63, all predator-prey models) under sigmoid function.
Interestingly, some cycle length is as long as 8 (8C).

{\bf Situations not covered:}  Besides a small positive or negative value added
to the threshold to create new variants (e.g. Eq.(\ref{m1p1-pos}) and Eq.(\ref{m1p1-neg})), 
the threshold is not treated as a new parameter. The reason is that we treat
each MP model as a point on a grid or hypercube. It is not appropriate to use larger
integer values as threshold as the node variable itself is limited to $[-1,1]$
or [0,1]. The choice of +1 or $-1$ is also of limited use. Instead, we use
a small non-zero threshold to study its effect by examining different variants.

Just because our discretized MPN models are on a grid in the model space, 
a smallest mutation is meant to be one Hamming distance in the integer space,
instead of a small change in non-integer space. 
If the question is on whether
dynamics is changed with a small perturbation on the parameter (e.g. 
$a \rightarrow a+\delta$ with $|\delta| \ll 1$), the resulting model is
no longer on the grid, and therefore, beyond the scope of this work.

One question in complex systems and networks, with a high level of importance,
is whether the limiting
dynamics of a network with a given logical structure and given variant
can be decided without simulation.  The related issue is relationship 
between a circuit's structure and its function \citep{jimenez17},
or reversely, how to find logical structure of a network with a given function/dynamics
\citep{tang21} (whether such a task is even feasible \citep{rocha22}),
with the latter having crucial consequences in synthetic biology 
\citep{chakraborty}. 
In this work, even if we design a regulatory graph  
by a MPN model, the constructed global interaction graph may have a
a different graph structure. Such a construction is equivalent to
running a simulation in one step.  The relationship between the
originally given and constructed graph structures and limiting dynamics,
needs further investigation.

There are many potential extensions of this work.
One obvious future study is to consider the case of MPNs with three
or more nodes. The number of nodes in a feedback loop
do play a role in determining the limiting dynamics \citep{remy}.
Certainly, we expect to obtain a richer dynamics 
such as higher cycle lengths in the limiting state.  
Besides the feed-forward loop regulation mentioned in Introduction
which requires at least three nodes, the exclusive-OR using MP gates
requires two more hidden nodes, besides the two input nodes and one output node.
The phenomenon of chimera state, where the identical and identically linked
nodes exhibit two different dynamics at different nodes
\citep{kuramoto,abrams,astero}, 
are only possible with many nodes and possibly nonlocal interactions and 
are thus  absent in the minimal $N=2$ MPNs.

\section*{Acknowledgements}

We would like to thank Yannis Almirantis for discussions on Rene Thomas'
work. W.L. would like to thank Hans-Peter Stricker for discussions
on forced direct graphs and folded spaces.

\newpage

\huge

{\bf Supplementary Material}

\large

\setcounter{figure}{0}
\renewcommand{\thefigure}{S\arabic{figure}}

\setcounter{equation}{0}
\renewcommand{\theequation}{S\arabic{equation}}

\setcounter{table}{0}
\renewcommand{\thetable}{S\arabic{table}}

\setcounter{section}{0}
\renewcommand{\thesection}{item-\arabic{section}}

\section{One-node (N=1) and zero-node (N=0) networks rules}
\label{sec:N0N1}

\indent

When a MPN rule has $b=c=0$, it is in fact a one-node rule.
These include Rule 13, 14, 15,  42, and 69. The Rule 41
has $a=b=c=d=0$, and it is a doing-nothing zero-node rule.

\begin{table}[H] \footnotesize
\begin{center}
\begin{tabular}{c|c|ccc|cc|ccc}
\hline
rule & links & \multicolumn{3}{c|}{equivalentR} &\multicolumn{2}{c|}{bipolar variants} & \multicolumn{3}{c}{ binary variants}  \\
R & $a$ $b$ $c$ $d$ & T12 & $Z_2$ & T12+$Z_2$& V1 & V2=V3 & V4=V7& V5 & V6 \\
\hline
13 & -1 0 0 -1 & 13 & 13 & 13 &  2C & 2C & F$_1$ & 2C & F$_1$ \\
14 & -1 0 0 0 & 40 & 14 & 40 & 2C & 2C &  F$_2$ & 2C & F$_1$\\
15 & -1 0 0 1 & 67 & 15 & 67 &  2C & 2C & F$_2$ & 2C & F$_2$ \\
41 & 0 0 0 0 &  41 & 41 & 41 & F$_4$ & F$_1$ & F$_4$ & F$_1$ & F$_1$\\
42 & 0 0 0 1 & 68 & 42 & 68  & F$_4$ & F$_2$ & F$_4$ & F$_1$ & F$_2$ \\
69 & 1 0 0 1 & 69 & 69 & 69 &  F$_4$ & F$_4$ & F$_4$ & F$_1$ & F$_4$\\
\hline
\end{tabular}
\end{center}
\caption{\label{tableS1-N1}
rule(R): rule number;
links $a,b,c,d$: four link parameters;
equivalent $R$ by T12, $Z_2$, T12+$Z_2$: rule numbers of those models
that are equivalent by exchanging two nodes, by discrete gauge transformation,
and by the combination of both;
bipolar variants V1, V2=V3: limiting dynamics for bipolar
($[-1, 1]$) variants V1 and V2 (V3 has identical dynamics as V2);
binary variants V4=V7, V5, V6: limiting dynamics for binary
([0,1]) variants V4 (V7 has identical dynamics as V4), V5, and V6.
}
\end{table}

\section{Three spaces: gene space, rule space, and network state space}
\label{Msec:spaces}

\indent

There are $N (N=2$) nodes in the gene space representing $N$ genes,
and directed links between them specify the nature of regulation.
In the fully discretized MPN rule space, there are $3^{N \times N}$ 
(with two nodes, $N=2$, $3^4=81$) nodes representing all possible network models/rules. 
Two rules are one Hamming distance away from each other
if one of the $a, b, c, d$ values in these two rules differ by 1,
while the value of the other parameters are kept the same.

In the network state space, there are $2^N$ (with two nodes, $N=2$, $2^2=4$)
points representing all possible $N$-gene network states.
These four network (or two-node) states can be
conveniently represented by four corners of a square. The point
at the origin is the $(-1,-1)$ state (for bipolar) or (0,0) state
(for binary). Moving along the $x$-axis to another corner
would be $(1,-1)$ for bipolar and (1,0) for binary. etc.
These four states can also be presented by four numbers, 0,1,2,3,
with (for bipolar) $(-1,-1)$ to be 0, $(1,-1)$ to be 2, etc.
The dynamics of updating of the state,
$(x^t, y^t) \rightarrow (x^{t+1}, y^{t+1})$,
introduces a link (arrow) from the first state and the second state.
Different types of dynamics in the state space will
be described in section \ref{Msec:limdyn}.

\section{New ``binary difference" variant V7 and its equivalent to V4}

\indent

Variants considered so far (V1-V6) are somewhat not satisfactory conceptually.
Bipolar variants assume a negative variable value which is not realistic
in most applications such as neuron potentials. On the other hand, the
positive or negative impact of one node on another is more reasonable
for changes, or differences, of node variable.
Take these two considerations,
we define a new model variant (V7) starting from:
\begin{eqnarray}
x^{t+1}- x^t &=& Sign(ax^t+by^t) \nonumber \\
y^{t+1}- y^t &=& Sign(cx^t+dy^t), \nonumber
\end{eqnarray}
then we move $x^t,y^t$ to the right-hand-side, and impose a step function
to force the mapping to [0,1] (assuming Sign(0)=0 and Step(0)=0):
\begin{eqnarray}
\label{eq-v7}
x^{t+1}  &=& Step(x^t+Sign(ax^t+by^t)) \nonumber \\
y^{t+1}  &=& Step(y^t+Sign(cx^t+dy^t))
\end{eqnarray}
In other words, we want to keep the possibility of $-1$
in $(a,b,c,d)$ parameter, but use only [0,1] in the node variable.
The V7 variant turns out to be equivalent to the V4 variant (binary as it is)
by the following argument.
If $x^t=0$, the Eq.(\ref{eq-v7}) and Eq.7 (in the main text)  lead to the same
outcome. If $x^t=1$, when $ax^t+by^t=0$, for V4, due to ``as it is" rule,
$x^{t+1}=x^t=1$, same as V7; when $ax^t+by^t=1$ (or $-1$), V4 leads
to $x^{t+1}=1$ (or 0), and for V7, Step(1+Sign(1))=1 (Step(1+Sign(-1))=Step(0)=0),
so the two have the same result.

\section{Shorthand notations of limiting dynamics}
\label{Msec:limdyn}

\indent

With $x$ and $y$ each taking only two values, the node state $(x,y)$
can only take four possible values. The number of possible dynamical
behaviors is also limited. We count all possible limiting dynamics
here and give each a shorthand notations.

\begin{itemize}
\item
F$_1$: all four initial states are attracted to one fixed-point
limiting dynamics. The eigenvalues of the corresponding state transition
matrix (after transpose) are (1,0,0,0).

\item
F$_2$: two fixed-point attractors, and two states are in the
(transient) basin of attractors. The related eigenvalues
are (1,1,0,0).

\item
F$_3$: three fixed-point attractors, and one state is in the
(transient) basin of attractors. The related eigenvalues are
(1,1,1,0).

\item
F$_4$: four fixed-point attractors, with the eigenvalues being
(1,1,1,1).

\item
2C: either single two-cycle attractor, with the eigenvalues being
$(1,-1,0,0)$, or two two-cycle attractors, with eigenvalues being
$(1,-1,1,-1)$.

\item
M (mixed):
either one two-cycle attractor and two fixed point attractors
(i.e. 2C+F$_2$) with $(1,-1,1,1)$ eigenvalues, or,
one two-cycle attractor and one fixed point attractor
(i.e. 2C+F$_1$) with $(1,-1,1,0)$.

\item
3C: one three-cycle attractor, with eigenvalues of
(1, $e^{2 \pi i/3}, e^{- 2 \pi i/3}$, 0).

\item
4C: one four-cycle attractor, with eigenvalues of
$(1,i, -1, -i)$.
\end{itemize}
In principle, one can also have 3C+F$_1$ with eigenvalues
of (1,1, $e^{2 \pi i/3}, e^{- 2 \pi i/3}$ ), but we
have not found an example of it yet.

\section{Node name conventions}
\label{Msec:nodename}

\indent

In commensalism models, one of the nodes receives positive contribution
from another node without a return (e.g., $x$ benefits from $y$, $b=1$, $c=0$).
The node receiving benefit (e.g., $x$) is called ``commensal", and the
node providing the benefit (e.g., $y$) called ``host".
In amensalism models, the node (e.g., $x$, if $b=-1$) receiving negative
contribution from another node (e.g., $y$) without a return is called
``victim", and the node imposing the action is called ``dominator".
For both unidirectional logical structures, we call the node
providing the link ``donor" and the node receiving the link ``receiver".
For predator-prey models, the node receiving the negative action
is ``prey" and the node receiving the benefit is ``predator".
For symmetric models (mutualism, competition), there is no need to
distinguish the two nodes.

If a MPN model has self-pointing loop, if the sign is positive,
we call the self-loop ``autocatalysis" (autocatalytic);
if the sign is negative, the self-loop is ``self-regulation" (self-regulating).

\section{Discrete gauge transformation $Z_2$ 
in ``bipolar as it is" variant (V1)}
\label{Msec:gauge}

\indent

The V1 variant (bipolar as it is) has an extra equivalence  transformation
that will further reduce the number of equivalent models: the 
discrete gauge transformation (see \citep{toulouse77,toulouse80} and 
the supplement material of \citep{wagner-martin07-plos}).
If both $b$ and $c$ change sign, we have the updating mapping as
$x^{t+1}= Sign( a x^t -b y^t)$ and $y^{t+1} = Sign (-c x^t +d y^t)$.
If we define a new variable $x'=-x$, it can be seen that the updating equation
for $(x', y)$ variable is exactly the same as the original equation.

For V2 variant, when $b$ and $c$ change sign, we have
$x^{t+1}= Sign( a x^t -b y^t +\epsilon)$ and $y^{t+1} = Sign (-c x^t +d y^t+\epsilon)$.
For a new variable $x'=-x$, although the updating function for $y$
is unchanged, the updating function for $x'$ is changed to
$x^{t+1}= Sign( a x^t +b y^t -\epsilon)$, with a different threshold.

\section{Transformation between bipolar and binary variant models}

\indent

The difference between bipolar  models and binary  models can be explained
by this variable transformation ($x,y \in [-1,1]$, $x',y' \in [0,1]$):
\begin{eqnarray}
  x' &=& \frac{x+1}{2} \nonumber \\
  y' &=& \frac{y+1}{2}
\end{eqnarray}
and V1 (Eq.2 in the main text)  can be written as:
\begin{eqnarray}
x'^{t+1}  & = & \left\{
\begin{array}{cl}
\text{Step} \left( a' x'^t +b' y'^t - \frac{a'+b'}{2} \right)  &
 \text{if} \hspace{0.1in} a'x'^t+b'y'^t \ne  (a'+b')/2    \\
x'^t & \text{if} \hspace{0.1in}  a'x'^t+b'y'^t= (a'+b')/2
\end{array}
\right.  \nonumber \\
y'^{t+1}  &=& \left\{
\begin{array}{cl}
\text{Step} \left( c' x'^t +d' y'^t - \frac{c'+d'}{2} \right)  &
 \text{if} \hspace{0.1in} c'x'^t+d'y'^t \ne  (c'+d')/2    \\
y'^t & \text{if} \hspace{0.1in}  c'x'^t+d'y'^t= (c'+d')/2
\end{array}
\right.
\end{eqnarray}
where $a'=2a$, $b'=2b$, $c'=2c$, and $d'=2d$. Not only the parameter grid
of $(a,b,c,d) \in [-1,0,1]$ is very different from that
for $(a',b',c',d') \in [-1,0,1]$, but also the threshold is set at a potential
non-zero point ($(a'+b')/2$ and $(c'+d')/2$).

\section{The original regulatory graph and the constructed ``global interaction graph",
in the gene/node space}
\label{Msec:intgraph}

\indent

When the four parameters $(a,b,c,d)$ in a MPN model are given,
we can draw a graph showing the impact of one node to another,
or to itself, is called (signed) regulatory graph in this paper.
Examples include those in Fig.1(B) and Fig.2 (in the main text).
On the other hand, in many other situations, such graph is not
available  when the mapping function is defined as complicated
Boolean/logic functions/gates. In order to determine the impact of one node
on another, an algorithm was proposed to construct 
the ``global interaction graphs" \citep{richard}.

This algorithm as applied to $N=2$ MPNs is the following:
we follow 4 paths along the edge of a square, two horizontal
arrows along the $x$ direction ((e.g., for binary) (0,0) to
(1,0), and (0,1) to (1,1)), and two vertical arrows along the
$y$ direction ((0,0) to (0,1), and (1,0) to (1,1)). The first two
actions increase the $x$ value by one unit, and the next two actions
increase the $y$ value by one unit.

Suppose a model maps (0,0) to (1,0), and maps (1,0) to (1,1),
the first bit of the two (for $x$) mapped network states is the same (1=1),
but the second bit (for $y$) is increased by one unit (0 to 1). It implies
that an action of increasing $x$ lead to no change in $x$
(there is no self loop at $x$), but increases the value of $y$
(therefore, there is a positive link from $x$ to $y$).
When all four arrows are examined in a similar way, we can construct
an interaction graph with signs, called global interaction graph in \citep{richard2}.

\section{Sixteen basic Boolean logic gates}
\label{sec:16gate}

\indent

With two Boolean inputs and one Boolean output,
there are only 16 possible functions (gates). The names of these
16 gates are listed in Table \ref{tableS2-16gates}.

\begin{table}[H]
\begin{center}
\begin{tabular}{c|cccc|c}
\hline
gate &  \multicolumn{4}{c|}{output when $x,y=$} & effective num \\
name & 0,0 & 0,1 & 1,0 & 1,1 & of input\\
\hline
FALSE & 0 & 0 & 0 & 0 & 0\\
AND & 0 & 0 & 0 & 1  & 2 \\
$x$ AND $\bar{y}$ & 0 & 0 &1 & 0 & 2\\
$x$ & 0 & 0 & 1 & 1  & 1\\
$\bar{x}$ AND $y$ &0 &1 &0 &0 &2\\
$y$ & 0 &1 &0 &1  & 1\\
XOR & 0 & 1 & 1 & 0  & 2\\
OR & 0& 1&1 &1  & 2\\
NOR &1 &0 &0 &0  & 2 \\
NXOR &1 &0 &0 &1  & 2\\ 
$\bar{y}$ &1 &0 &1 &0 & 1 \\
$y$-imply &1 & 0& 1& 1 & 2\\
$\bar{x}$ & 1&1 &0 &0  & 1\\
$x$-imply &1 &1 &0 &1 & 2\\
NAND &1  & 1 & 1 & 0  & 2\\
TRUE & 1& 1& 1& 1 & 0\\
\hline
\end{tabular}
\end{center}
\caption{\label{tableS2-16gates}
List of all logical gates with two inputs $x,y$.
The corresponding outputs of a gate with the four possible
input values are listed (e.g., output values when  $x=y=0$
are listed in the second column). The overhead bar means
logic NOT. The effective number of inputs
in the last column indicates the level of canalizing
(item-10). 
}
\end{table}

\section{Models in V1 (bipolar as it is) variant are all single-input gates}
\label{Rsec:V1}

\indent

In the context of Boolean functions or logic gates, a canalizing function has the property that 
one of its variables can determine the output, regardless of the state of the 
other variables \citep{kauffman-sa2}.  In biological systems, canalizing functions are significant 
because they model how certain genes or regulatory elements can have a dominant 
influence on the behavior of a gene network. Canalizing functions are essential 
for the robustness and stability that characterize biological systems.

There are 16 possible two-input-one-output Boolean functions (see item-9). 
These are:
FALSE, AND, $x$ AND $\bar{y}$, $x$, $\bar{x}$ AND $y$, $y$, XOR, OR, NOR, NXOR, 
$\bar{y}$, $y$-imply, $\bar{x}$, $x$-imply, NAND, TRUE.  The bar 
above the symbols denotes negation, equivalent to NOT; NOR can also be 
written as $\overline{OR}$, NXOR as $\overline{XOR}$, NAND as $\overline{AND}$.
For bipolar models, replace 0 in Table \ref{tableS2-16gates} by $-1$.
Some of these Boolean functions are essentially zero-input (FALSE, TRUE), 
or one-input ($x, y, \bar{x}, \bar{y}$), thus canalizing functions. 
The imply-$x$ and imply-$y$ might be considered as partially  canalizing
\citep{reichhardt} as it becomes one-input function conditional on the second input.

Table \ref{tableS3-21V1} shows the equivalent logic gates of 21 equivalent
MPN models in bipolar as it is variant (V1).
We arrange the models so that those with the same dynamics are next
to each other.  By examining the 21 representative rules in Table \ref{tableS3-21V1}
(as well as 18 other rules that are equivalent to these rules by
discrete gauge transformation), we found that all of mapping rules can be rewritten 
as single-input Boolean functions. For F$_4$ models, $x'=x$, $y'=y$, which is
expected as each state maps to itself. For F$_2$ rules, $x$ becomes
a copy (with negation) of the $y$, and $y$ maps to itself. The two-node/gene
system essentially follows a single-node/gene dynamics. 

For two-cycle models (2C), one node/gene negates itself, and the second 
node/gene either maps to itself or becomes redundant by being a copy 
of the first node/gene.
The four-cycle models (4C), $x'=\overline{y}, y'=x$,
can be viewed as a reflection with respect to a horizontal, followed
by a switch between $x$ and $y$ axes.

\begin{table}[H] \normalsize
\begin{center}
\begin{tabular}{c|c|c|cc|c}
\hline
V1 dynamics & rule & a b c d &  $Z_2$ & T12+$Z_2$ &  V1 logic gates  \\
\hline
$F_4$ & 39 & 0 0 -1 1 &   45 & 77 & $x'=x, y'=y$    \\
& 60 & 1 -1 0 1 &  78 & 72 &  same  \\ 
& 57 & 1 -1 -1 1 &   81 & 81 &  same   \\
& 63 & 1 -1 1 1 &  75 &63 &  same  \\ 
& \multicolumn{2}{l}{45} &  &  &  same \\
& \multicolumn{2}{l}{72,81} &  &  &   same  \\
\hline
$F_2$ & 3 & -1 -1 -1 1 &   27 & 79 &   $x'$=NOT($y$), $y'=y$ \\
& 30 & 0 -1 -1 1 & 54 &80 &   same  \\
& 9 & -1 -1 1 1 &  21 & 61 &    same \\
& 36 & 0 -1 1 1 &  48 &62 &    same  \\
& 5 & -1 -1 0 0 & 23 & 43 &    same  \\
& 6 & -1 -1 0 1 &  24 & 70 &    same  \\
& 32 & 0 -1 0 0 & 50 &44 &    same   \\
& 33 & 0 -1 0 1 & 51 &71 &   same   \\
& \multicolumn{2}{l}{23,24,51}&   & &  $x'=y,y'=y$  \\
& \multicolumn{2}{l}{21,48}&    & &   same \\
& \multicolumn{2}{l}{27,54}&    & &   same \\
& \multicolumn{2}{l}{44}&   & &   $x'=x,y'=x$  \\
\hline
$M$ & 1 & -1  -1 -1 -1 &    25 & 25  & $x'$=NOT($y$), $y'$=NOT($x$)  \\ 
& 2 & -1  -1 -1 0 &  26 & 52   & same  \\
& 29 & 0  -1 -1 0 &  53 &53 & same   \\
& \multicolumn{2}{l}{25,26,53} &  &  & $x'=y,y'=x$ \\
\hline
2C & 4 & -1 -1  0 -1 &   22 & 16 &   $x'$=NOT($y$), $y'$=NOT($y$)  \\
& 11 &  -1 0 -1 0 &  17 & 49 &    $x'$=NOT($x$), $y'$=NOT($x$) \\
& 12 &  -1 0 -1 1 &  18 &76 & $x'$=NOT($x$), $y'=y$   \\
 & \multicolumn{2}{l}{18 }&   &   & same   \\
 & \multicolumn{2}{l}{16,17 } &  &  &  $x'$=NOT($x$),$y'=x$  \\
\hline
4C & 7 & -1  -1 1 -1 &   19 & 7 &     $x'$=NOT($y$), $y'=x$   \\
 & 8 & -1   -1 1 0 &   20 & 34 &  same   \\
 & 35 & 0  -1 1 0 &   47 &35 &   same   \\
 & \multicolumn{2}{l}{20 } &  & &  $x'=y, y'$=NOT($x$)   \\
\hline
\end{tabular}
\end{center}
\caption{ \label{tableS3-21V1}
The list of 21 MPN models organized by their V1 limiting dynamics,
in the F$_4$, F$_2$, M, 2C, 4C order,
i.e., models with the most number of fixed-points at the top,
and models with the longest cycle length at the bottom.
The discrete-gauge-transformation ($Z_2$) equivalent rules, though have the same
limiting dynamics,  may or may not have the same logic gate.
If not, the logic gate is also listed. 
The equivalent logic gate of each model in V1 is described in the last column.
}
\end{table}

\section{Other variants can be two-input gates thus are less canalizing}
\label{sec:logic}

\indent

Previously,  we conclude that all 21 V1 rules contain only 
single-gene canalizing logic gates. 
In that case, even if there are two inputs to a node, its next time value is 
completely determined by only one input.  However, V1 variant is an exception: 
other MPN variants show a whole range of logic gates.

Table \ref{tableS4-logic} shows the equivalent logic gates for all 39 rules
in variants V1-V6. Although V2 and V3 exhibit the same limiting
dynamics, the equivalent logic gates can be different. For example, for
R32, V2 updates the state by: $x^{t+1}=$NOT$(y^t)$ and $y^{t+1}=$TRUE (always 1),
whereas V3 updates by $x^{t+1}=$NOT$(y^t)$ and $y^{t+1}=$FALSE  (always 0).

Comparing Table \ref{tableS4-logic} and Table \ref{tableS3-21V1}, one may see which
logic gates lead to what types of limiting dynamics. For example, if $x$ and/or $y$ 
updates as TRUE (1) or FALSE (-1 or 0), the limiting dynamics is F$_1$. If $x$ and $y$
switch places (e.g., $x^{t+1}= y, y^{t+1}=x$), there will be cycles
(which could co-exist with fixed-point as a M type dynamics). As V1 always
contains most diverse types of dynamics, it is interesting that more
complicated logic gates (less canalizing) doesn't necessarily make the
dynamics more complex or diverse.

\begin{table}[H] \footnotesize
\begin{center}
\begin{tabular}{c|ccc|ccc}
\hline
rule & \multicolumn{3}{c|}{ bipolar rules} & \multicolumn{3}{c}{ binary rules}\\
\cline{2-7}
R & V1 & V2 & V3  &  V4 & V5 &   V6 \\
\hline
1 & $\bar{y},\bar{x}$ & NAND,NAND & NOR,NOR &  F,F & NOR.NOR & F,F  \\
2 & $\bar{y},\bar{x}$ & NAND,$\bar{x}$ & NOR,$\bar{x}$ &    F,$\bar{x}$ANDy & NOR,$\bar{x}$ & F,F   \\
3  & $\bar{y}$,$y$ & NAND,$x$IMP & NOR,$\bar{x}$AND$y$ &   F,$y$ & NOR,$x$IMP & F,$\bar{x}$AND$y$   \\
4  & $\bar{y}$,$\bar{y}$ & NAND,$\bar{y}$ & NOR,$\bar{y}$   & F,F & NOR,$\bar{y}$ & F,F   \\
5  & $\bar{y}$, $y$ & NAND,T & NOR,F  &  F, $y$ & NOR,T & F,F   \\
6  & $\bar{y}$, $y$ & NAND, $y$ & NOR, $y$  &  F, $y$ & NOR,T & F, $y$  \\
7  & $\bar{y}$, $x$ & NAND,$y$IMP & NOR,$x$AND$\bar{y}$   & F, $x$ & NOR,$y$IMP & F,$x$AND$\bar{y}$  \\
8  & $\bar{y}$, $x$ & NAND, $x$ & NOR, $x$  &  F,OR & NOR,T & F, $x$  \\
9  & $\bar{y}$, $y$ & NAND,OR & NOR,AND & F,OR & NOR,T & F,OR \\
11  & $\bar{x}$,$\bar{x}$ & $\bar{x}$,$\bar{x}$ & $\bar{x}$,$\bar{x}$  & F,$\bar{x}$AND $y$ & $\bar{x}$,$\bar{x}$ & F,F   \\
12  & $\bar{x}$, $y$ & $\bar{x}$,$x$IMP & $\bar{x}$,$\bar{x}$AND $y$   & F, $y$ & $\bar{x}$,$x$IMP & F,$\bar{x}$AND $y$ \\
16  & $\bar{x}$, $x$ & $\bar{x}$,$y$IMP & $\bar{x}$,$x$AND$\bar{y}$  & F, $x$ & $\bar{x}$,$y$IMP & F,$x$AND$\bar{y}$ \\
17  & $\bar{x}$, $x$ & $\bar{x}$, $x$ & $\bar{x}$, $x$    & F,OR & $\bar{x}$,T & F, $x$  \\
18  & $\bar{x}$, $y$ & $\bar{x}$,OR & $\bar{x}$,AND    & F,OR & $\bar{x}$,T & F,OR  \\
20  & $y$,$\bar{x}$ & $x$IMP,$\bar{x}$ & $\bar{x}$AND$y$,$\bar{x}$  &  $y$,$\bar{x}$AND $y$ & $x$IMP,$\bar{x}$ & $\bar{x}$AND$y$,F \\
21  & $y$, $y$ & $x$IMP,$x$IMP & $\bar{x}$AND$y$,$\bar{x}$AND $y$   &  $y$, $y$ & $x$IMP,$x$IMP & $\bar{x}$AND$y$,$\bar{x}$AND $y$  \\
23  & $y$, $y$ & $x$IMP,T & $\bar{x}$AND$y$,F   &   $y$, $y$ & $x$IMP,T & $\bar{x}$AND$y$,F  \\
24  & $y$, $y$ & $x$IMP, $y$ & $\bar{x}$AND$y$, $y$   &  $y$, $y$ & $x$IMP,T & $\bar{x}$AND$y$, $y$  \\
25  & $y$, $x$ & $x$IMP,$y$IMP & $\bar{x}$AND$y$,$x$AND$\bar{y}$   & $y$, $x$ & $x$IMP,$y$IMP & $\bar{x}$AND$y$,$x$AND$\bar{y}$   \\
26  & $y$, $x$ & $x$IMP, $x$ & $\bar{x}$AND$y$, $x$  & $y$,OR & $x$IMP,T & $\bar{x}$AND$y$, $x$   \\
27  & $y$, $y$ & $x$IMP,OR & $\bar{x}$AND$y$,AND    & $y$,OR & $x$IMP,T & $\bar{x}$AND$y$,OR  \\
29  & $\bar{y}$,$\bar{x}$ & $\bar{y}$,$\bar{x}$ & $\bar{y}$,$\bar{x}$   & $x$AND$ \bar{y}$,$\bar{x}$ANDy& $\bar{y}$,$\bar{x}$ & F,F   \\
30  & $\bar{y}$, $y$ & $\bar{y}$,$x$IMP & $\bar{y}$,$\bar{x}$AND $y$  & $x$AND$ \bar{y}$, $y$ & $\bar{y}$,$x$IMP & F,$\bar{x}$AND $y$   \\
32  & $\bar{y}$, $y$ & $\bar{y}$,T & $\bar{y}$,F   & $x$AND$ \bar{y}$, $y$ & $\bar{y}$,T & F,F  \\
33  & $\bar{y}$, $y$ & $\bar{y}$, $y$ & $\bar{y}$, $y$   & $x$AND$ \bar{y}$, $y$ & $\bar{y}$,T & F, $y$   \\
35  & $\bar{y}$, $x$ & $\bar{y}$, $x$ & $\bar{y}$, $x$  &  $x$AND$ \bar{y}$,OR & $\bar{y}$,T & F, $x$   \\
36  & $\bar{y}$, $y$ & $\bar{y}$,OR & $\bar{y}$,AND  &  $x$AND$ \bar{y}$,OR & $\bar{y}$,T & F,OR  \\
39  & $x$, $y$ & T,$x$IMP & F,$\bar{x}$AND $y$   & T,$x$IMP & $x$, $y$ & F,$\bar{x}$AND $y$  \\
44  & $x$, $x$ & T, $x$ & F, $x$  &  $x$,OR & T,T & F, $x$  \\
45  & $x$, $y$ & T,OR & F,AND  & $x$,OR & T,T & F,OR  \\
48  & $y$, $y$ & $y$,$x$IMP & $y$,$\bar{x}$AND $y$ &   OR, $y$ & T,$x$IMP & $y$,$\bar{x}$AND $y$   \\
51  & $y$, $y$ & $y$, $y$ & $y$, $y$  &  OR, $y$ & T,T & $y$, $y$   \\
53  & $y$, $x$ & $y$, $x$ & $y$, $x$  & OR,OR & T,T & $y$, $x$   \\
54  & $y$, $y$ & $y$,OR & $y$,AND &  OR,OR & T,T & $y$,OR \\
57  & $x$, $y$ & $y$IMP,$x$IMP & $x$AND$ \bar{y}$,$\bar{x}$AND $y$  & $x$, $y$ & $y$IMP,$x$IMP & $x$AND$\bar{y}$,$\bar{x}$AND $y$    \\
60  & $x$, $y$ & $y$IMP, $y$ & $x$AND$\bar{y}$, $y$   & $x$, $y$ & $y$IMP,T & $x$AND$\bar{y}$, $y$   \\
63  & $x$, $y$ & $y$IMP,OR & $x$AND$\bar{y}$,AND &  $x$,OR & $y$IMP,T & $x$AND$\bar{y}$,OR    \\
72  & $x$, $y$ & $x$,OR & $x$,AND & $x$,OR & T,T & $x$,OR  \\
81  & $x$, $y$ & OR,OR & AND,AND & OR,OR & T,T & OR,OR  \\
\hline
\end{tabular}
\end{center}
\caption{ \label{tableS4-logic}
The logic gate representation for 39 MPN rules for variants V1, $\cdots$ V6. 
For example, ``$x$IMP,OR" means
$x^{t+1}= x^t$ IMPLY (see Supplement Table S2), $y^{t+1}= $x$ $ OR $y$.
Those for V1 are already presented in Table \ref{tableS3-21V1}.
}
\end{table}

\vspace{0.5in}

\large

\end{document}